\documentclass[12pt]{iopart}
\usepackage{amsfonts,amssymb,graphicx}

\begin{document}

\title{Spectral statistics of a pseudo-integrable map: the general case}
\author{E. Bogomolny, R. Dubertrand$^{*}$, and \fbox{ C. Schmit}}
\address{\it Universit\'e Paris-Sud, CNRS,  UMR 8626\\
Laboratoire de Physique Th\'eorique et Model\`es Statistiques\\
91405 Orsay Cedex, France\\
$^{*}$Department of Mathematics\\
University of Bristol BS8 1TW, UK
}

\begin{abstract}
It is well established numerically  that spectral statistics of
pseudo-integrable models differs
 considerably from the reference statistics of integrable and chaotic
 systems. 
In [PRL {\bf 93} (2004) 254102] statistical properties of a certain quantized 
pseudo-integrable map had been calculated analytically but only for a special 
sequence of matrix dimensions. The purpose of this paper is to obtain the
spectral 
statistics of the same quantum  map for all  matrix dimensions.

\end{abstract}
\date{\today}
\maketitle

\section{Introduction}
The relation between spectral statistics of a quantum system and its classical
counterpart is one of the main achievements of quantum chaos. It is
established that at the scale of the mean level density the spectral
statistics of classically integrable systems are described by the Poisson
distribution \cite{Berry} and the spectral statistics of classically chaotic
systems are close to the statistics of eigenvalues of standard random matrix
ensembles depending only on the underlying symmetry \cite{Bohigas}. Though
these statements had not  mathematically been proved  in the full generality
and there exist noticeable exceptions (e.g. chaotic systems on constant
negative curvature surfaces generated by arithmetic groups \cite{Georgeot}),
these conjectures are widely accepted for 'generic' quantum systems. 

Much less is known when a system is not classically integrable or completely
chaotic. An important example which we have in mind is the case of
pseudo-integrable systems (see e.g. \cite{Richens}) represented by
2-dimensional polygonal billiards whose each angle is a rational multiple of
$\pi$. A typical classical trajectory in such models covers a 2-dimensional
surface of a finite genus $\geq 2$ \cite{katok}. For comparison, a trajectory
of a  2-dimensional integrable model belongs to a 2-dimensional torus of genus
1 and a typical trajectory of a 2-dimensional chaotic system covers
ergodically the whole 3-dimensional surface of constant energy.  
 
Numerical calculations \cite{BogomolnySchmit} clearly demonstrated  that the
spectral statistics of pseudo-integrable billiards is not universal and
depends on billiard angles. It appears that spectral statistics of such models
has two characteristic features: a level repulsion at small distances and an
exponential decrease of the nearest-neighbour distribution at large
separations. This type of statistics (called intermediate statistics)  has
been  observed for the first time in the numerical simulations  of the
Anderson model at the point of metal-insulator transition \cite{Shclovskii}
and later in this context has been profoundly  investigated (see e.g. a recent
review \cite{mirlin} and references therein).   
Unfortunately analytical progress in the investigation of intermediate
statistics in pseudo-integrable systems is limited. In \cite{Giraud} the level
compressibility of  certain pseudo-integrable billiards was computed
analytically, thus confirming the intermediate character of their spectral
statistics. The main difficulty in analytical treatment of 2-dimensional
pseudo-integrable billiards is the strong diffraction on  billiard corners
with angles $\neq \pi/n$ with integer $n$. Though there exists the exact
Sommerfeld's solution for the scattering on such angles \cite{Sommerfeld}, the
diffraction coefficient is formally infinite along the optical boundaries so
it is 
impossible to treat the multiple corner diffraction in perturbation series. A
certain progress has been made in \cite{superscars} where it was found that
the multiple diffraction along periodic orbit channels in pseudo-integrable
systems forces the wave functions to tend to zero on the channel boundaries thus
forming superscarring states observed recently in micro-wave experiments
\cite{Richter}. 

The difficulties in analytical solution of pseudo-integrable billiards lead to
the necessity of investigation of simpler models with similar features. A
promising example is the quantization of interval exchange maps which are
known to be the correct description of classical dynamics in pseudo-integrable
systems \cite{Richens}, \cite{katok}.  

In \cite{Marklof} the following classical parabolic  map had been quantized 
\begin{equation}
\Phi_{\alpha}: \left ( \begin{array}{c} p\\q\end{array}\right )\Longrightarrow
 \left ( \begin{array}{c} p+\alpha\\q+f(p+\alpha)\end{array}\right )\;
 \mbox{ mod }1\ .
\label{map}
\end{equation}
Here $\alpha$ is a constant and $f(p)$ is a certain function (taken equal $2p$
in \cite{Marklof}). For rational $\alpha=m/q$ this map corresponds to the
simplest interval exchange map of 2 intervals.  
 
After a straightforward generalization of the results of \cite{Marklof} the
$N\times N$ unitary matrix associated with the above map has the form of the
diagonal matrix $\mathrm{e}^{{\rm i}\Phi_k}$ with real $\Phi_k$ multiplied by a
constant unitary matrix $\mu_{kp}$ ($k,p=0,1,\ldots,N-1$) 
\begin{equation}
 M_{kp}={\rm e}^{{\rm i}\Phi_k}\mu_{kp}\;,
\label{matrix}
\end{equation}
where
\begin{equation}
 \mu_{kp}=\frac{1}{N}\sum_{r=0}^{N-1}\exp \left [\frac{2\pi {\rm i}}{N} r(k-p+\alpha N)\right ]=
\frac{1-{\rm e}^{2\pi{\rm i}\alpha N}}{N(1-{\rm e}^{2\pi{\rm i}(k-p+\alpha N)/N})}\;.
\label{mu}
\end{equation}
The matrix $\mu_{kp}$ depends on a parameter $\alpha$ and the last equality
are valid when $\alpha N$ is not an integer. (In the latter case the spectrum
of (\ref{matrix}) can be obtained analytically \cite{zeev} and we always
assume below that for rational $\alpha=m/q$, $mN\not \equiv  0$ mod $q$.)  

Two main cases are of interest. The first corresponds to the ensemble of
random matrices (\ref{matrix}) where all $N$ phases $\Phi_k$ are considered as
independent random variables distributed uniformly between $0$ and $2\pi$. We
call  such matrices  non-symmetric ensemble.  

In the second case only a half of the phases are independent  random variables uniformly distributed between $0$ and $2\pi$. The other half is related to the first one by the symmetry
\begin{equation}
 \Phi_{N-k}=\Phi_k\;, \;\;\;k=1,\ldots, \left [\frac{N}{2}\right ].
\label{symmetry}
\end{equation}
This case will be called symmetric ensemble as in the dynamical interpretation
the transformation $k\to-k$ is the time-inversion symmetry.  

\begin{figure}
\begin{minipage}{.49\linewidth}
\begin{center}
\includegraphics[width=.8\linewidth, angle=-90]{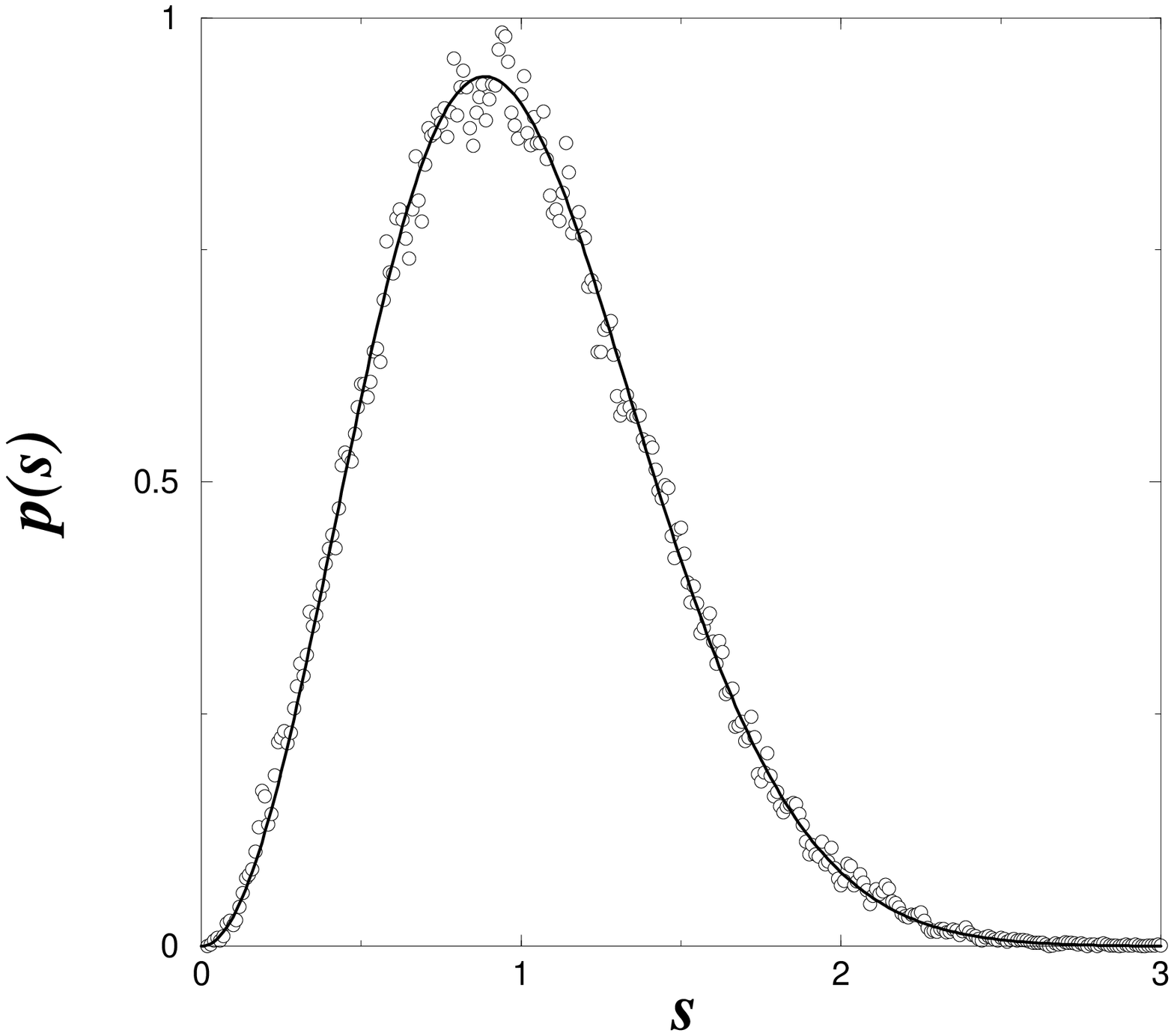}
\end{center}
\begin{center}a)\end{center}
\end{minipage}\hfill
\begin{minipage}{.49\linewidth}
\begin{center}
\includegraphics[width=.8\linewidth, angle=-90]{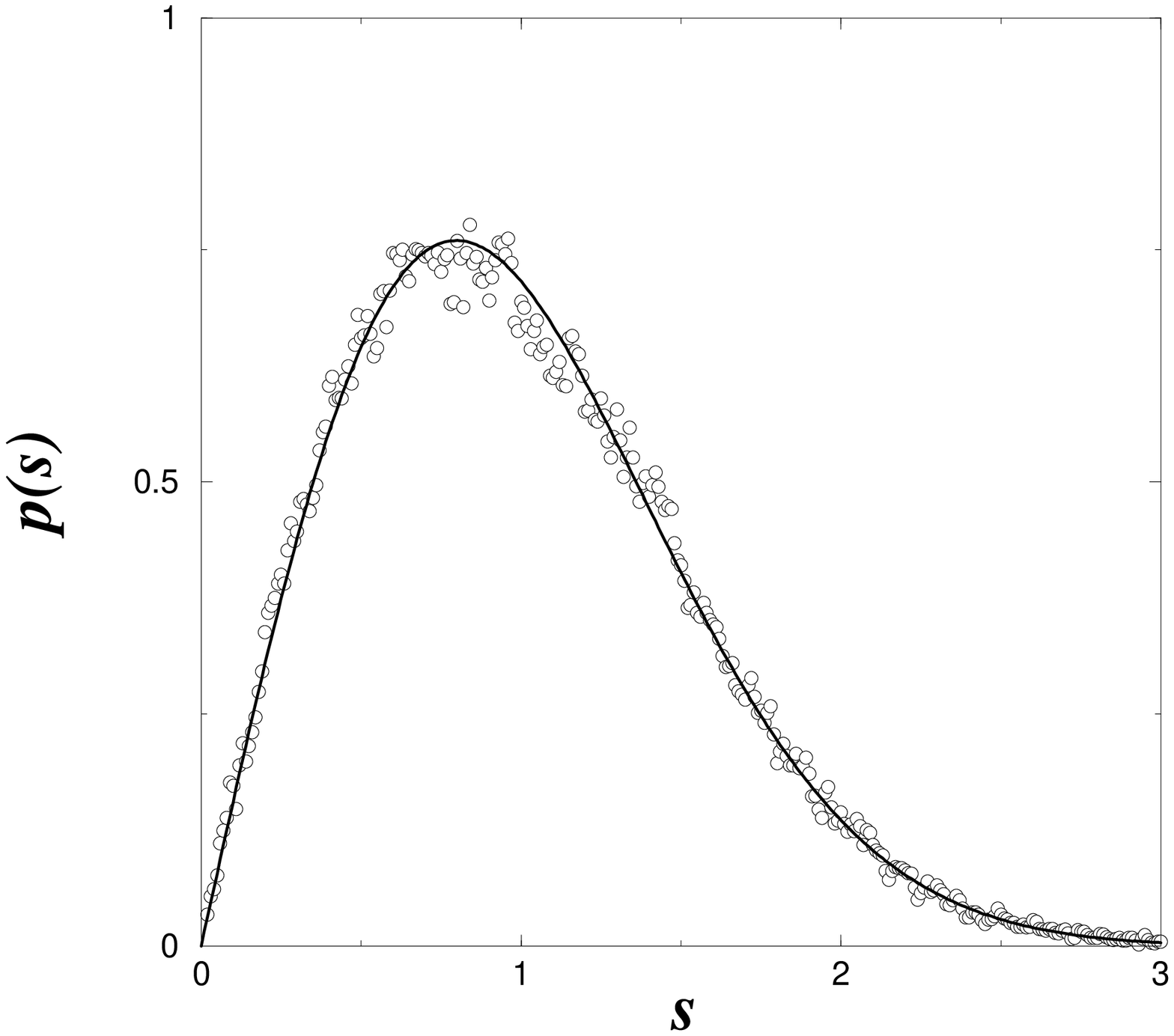}
\end{center}
\begin{center}b)\end{center}
\end{minipage}
\caption{Open circles are nearest-neighbour distribution with
  $\alpha=\sqrt{5}/4$ and $N=801$ for non-symmetric matrices  (a) and
  symmetric matrices (b). Solid lines indicate the Wigner surmise
  (\ref{surmise}). } 
\label{wiegner}
\end{figure}

The eigenvalues $\Lambda_n$ and the eigenvectors $u_k(n)$, $n=1,2,\ldots, N$,
are defined as usual 
\begin{equation}
 \Lambda_nu_k(n)=\sum_{p=0}^{N-1}M_{kp}u_p(n)\;.
\label{main}
\end{equation}
Because  matrix $M_{kp}$ is unitary ($M\,M^{\dag}=1$), its eigenvalues lie on the unit circle
\begin{equation}
 \Lambda_n={\rm e}^{{\rm i}\theta_n}
\end{equation}
and its eigenvectors can be chosen orthogonal
\begin{equation}
 \sum_{k=0}^{N-1}\bar{u}_k(m)u_k(n)=\delta_{mn} 
\label{orthogonal}
\end{equation}
The statistical properties of matrix (\ref{matrix}) depend crucially on
the arithmetic of the parameter $\alpha$. For irrational $\alpha$ the map
(\ref{map}) 
is only a parabolic map and methods developed in \cite{Schmit}  and in the
present paper can not be directly applied. Numerical calculations suggest that
for Diophantine $\alpha$ the spectral statistics of the matrix (\ref{matrix})
is very 
close to the standard statistics of the Gaussian ensembles of random matrices
\cite{mehta}. Namely, non-symmetric matrices are described by the GUE
statistics and symmetric matrices by the GOE statistics. For illustration in
Fig.~\ref{wiegner} we plot the nearest-neighbour distribution for matrices
(\ref{matrix}) with $\alpha=\sqrt{5}/4$ and $N=801$. The solid lines indicates
the Wigner surmise (\ref{surmise}) which is known (see e.g. \cite{mehta}) to
be a good approximation for the GUE and the GOE distributions.  
\begin{equation}
p_{\mathrm{GUE}}(s)=\frac{32}{\pi^2}s^2\mathrm{e}^{-4s^2/\pi}\;, \;\;
p_{\mathrm{GOE}}(s)=\frac{\pi}{2}s\mathrm{e}^{-\pi  s^2/4}\ . 
\label{surmise}
\end{equation} 
Rational $\alpha=m/q$ with co-prime   integers $m$ and $q$ correspond to an
interval exchange map \cite{Marklof} and we shall consider only such values of
$\alpha$ throughout the paper.  

This paper  investigates the following question: What are the statistical
properties of eigenvalues of the matrix (\ref{matrix}) for fixed rational
$\alpha=m/q$ and large $N$? It appears that to get a well defined limit in
this case it is necessary to consider increasing sequences of matrix
dimensions, $N$, such that the product of the numerator of $\alpha$ times $N$
has a fixed residue modulo the denominator of $\alpha$ 
\begin{equation}
mN\equiv k\;\;\mathrm{mod}\;\; q\ .
\label{modulus_k}
\end{equation} 
In \cite{Schmit} it was demonstrated that for a special sequence of matrix
dimensions with 
\begin{equation}
mN\equiv \pm 1\;\;\mathrm{mod}\;\; q
\label{modulus}
\end{equation}  
eigenvalues of the matrix (\ref{matrix}), (\ref{mu})  are described by the
so-called semi-Poisson statistics which has been proposed in \cite{Gerland} as
the simplest model of intermediate statistics.  

Let $x_1\leq x_2\leq \ldots \leq x_K$ be  an ordered sequence of real numbers
(eigenvalues). The joint distribution for the semi-Poisson statistics is
proportional to the product of the nearest distances between these level times
a confining potential $V(x)$ 
\begin{equation}
 P(x_1,x_2,\ldots,x_K)\sim \prod_{i}|x_{i+1}-x_i|^{\beta} \prod_i
 \mathrm{e}^{-V(x_i)} 
\end{equation}
In the limit $K\to \infty$ all correlation functions of the semi-Poisson
statistics at the scale of the mean level density do not depend on $V(x)$ and
can be obtained analytically \cite{Gerland}. In particular, the probability
that between 2 levels there exist exactly $n-1$ levels  has the form 
\begin{equation}
 p_n(\beta;s)=\frac{(\beta+1)^{n\beta+n}}{\Gamma(n\beta+n)} s^{n\beta+n-1}
 \mathrm{e}^{-(\beta+1)s}\ . 
\end{equation}
The semi-Poisson statistics depends only on one parameter $\beta$ which fixes
the level repulsion at small distances so the nearest-neighbour distribution
(i.e. $p_n(\beta;s)$ for $n=1$) tends to zero as $s^{\beta}$ 
 \begin{equation}
 p(\beta;s)=A_{\beta}s^{\beta}\mathrm{e}^{-(\beta+1)s}
\label{nnd}
\end{equation} 
with $A_{\beta}=(\beta+1)^{\beta+1}/\Gamma(\beta+1)$.

From a mathematical point of view the semi-Poisson statistics can be considered
as a  stochastic process with independent increments (with gamma-distribution)
forming a convolution semigroup  
\begin{equation}
(p_n*p_m)(\beta;s) \equiv \int_0^s
p_n(\beta;y)p_m(\beta;s-y)\mathrm{d}y=p_{n+m}(\beta;s)\ . 
\label{convolution}
\end{equation}  
According to \cite{Schmit}  when the condition (\ref{modulus}) is satisfied
the spectral statistics of the matrix (\ref{matrix}) tends for large $N$ to
the semi-Poisson distribution with the following integer and half integer
values of $\beta$ depending on the denominator of $\alpha=m/q$ and the
symmetry of the map 
\begin{equation}
 \beta=\left \{\begin{array}{cl}q-1 &\mbox{ for non-symmetric ensemble}\\
     \frac{1}{2}q-1 &\mbox{ for symmetric ensemble}\end{array}
 \right . .  
\label{beta}
\end{equation}
To compare numerical calculations with theoretical predictions it is often
more precise to consider instead of  the nearest-neighbour distribution,
$p(s)$, (as in Fig.~\ref{wiegner})  its integral 
\begin{equation}
N(s)\equiv \int_0^sp(s^{\prime})\mathrm{d}s^{\prime}
\label{integrated_nnd}
\end{equation}
which gives the relative number of levels when the distance between the
nearest-neighbour eigenvalues is less than $s$. This quantity is one of the
main spectral correlation functions and throughout the paper we focus
exclusively on it though other correlation functions can also be calculated
and are of interest. 

\begin{figure}
\begin{center}
\includegraphics[width=.4\linewidth, angle=270]{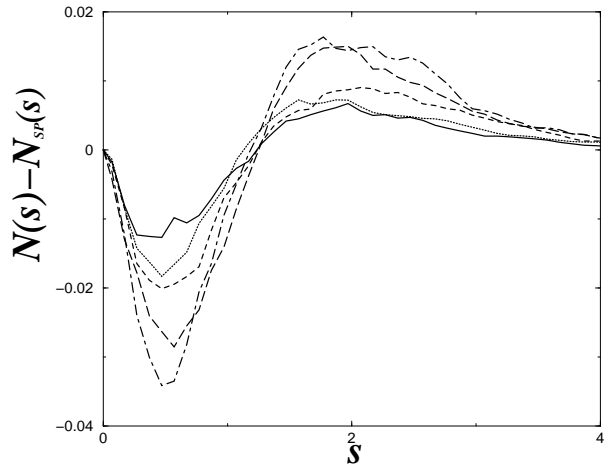}
\end{center}
\caption{Differences between the integrated nearest-neighbour distributions for
  the non-symmetric ensemble of matrices (\ref{matrix}) with $\alpha=1/2$ for
  different odd $N$ and the theoretical prediction for this case
  (\ref{theory}). The different lines  from bottom to top at small $s$
  correspond, respectively,  to $N=101$, $201$, $401$, $801$, and $1601$.} 
\label{fig0}
\end{figure}

To illustrate the convergence of the spectral statistics of the above matrices
to the predicted values let us consider e.g. $\alpha=1/2$ with odd $N$. From
(\ref{beta}) and (\ref{nnd})  it follows that the limiting integrated
nearest-neighbour distribution in the non-symmetric case is here the simplest
semi-Poisson distribution with $\beta=1$ 
\begin{equation}
N_{\mathrm{sp}}=1-(2s+1)\mathrm{e}^{-2s}\ . 
\label{theory}
\end{equation}
In Fig.~\ref{fig0} the difference between the integrated nearest-neighbour
distribution computed numerically  and its theoretical prediction
(\ref{theory})  is plotted for different odd matrix dimensions. For this and
other similar figures in the paper the number of realizations is chosen to be
the minimum between $100$ and $50000/N$. The figure  shows that the agreement
is quite good (less than $0.02$) even for $N$ of the order of a few hundreds.        

The purpose of this paper is to calculate the spectral statistics of the above
quantized pseudo-integrable map in the general case (\ref{modulus_k}) with $k
\neq 0,\pm 1$ mod $q$. The plan of the paper is the
following. Sections~\ref{rank_1} and \ref{long_range} give extended details of
the 
construction briefly discussed in \cite{Schmit}. The peculiarity of the
problem under consideration is the existence of 2 rank-one deformations of the
original matrix (\ref{matrix}) with known eigenvalues and
eigenvectors. These deformations are discussed in Section~\ref{rank_1}.
In Section~\ref{long_range} it is demonstrated that
these rank-one deformations lead to long-range correlations between
the eigenvalues of the initial matrix (\ref{matrix}). To obtain a clear
picture of 
these correlations it is convenient to use a special form of eigenvalue
ordering (an analog of the unfolding) which is discussed in
Section~\ref{geo_unfolding}. In Section~\ref{transfer} it is shown that these
long-range correlations can effectively be taken into account by the
construction 
of a kind of transfer operator. This operator is a finite dimensional matrix
whose largest eigenvalue and corresponding eigenvectors permit to calculate
all correlation functions. This is done explicitly for a few main examples in
Section~\ref{explicit}. Obtained analytical formulas agree well with numerical
calculations. The summary of the results is present in
Section~\ref{summary}. Certain technical details are given in~\ref{appendix}.   

\section{Rank-one deformations}\label{rank_1}

As was shown in \cite{Schmit} the important property of the matrix
(\ref{matrix}), (\ref{mu}) is the possibility to rewrite it in the following
form 
\begin{equation}
 M_{kp}\equiv \mathrm{e}^{{\mathrm{i}}\Phi_k}\frac{(1-{\rm e}^{2\pi {\rm
       i}\alpha N})}{N(1-{\rm e}^{2\pi {\rm i}(k-p+\alpha N)/N})}=
N_{kp}+\frac{1-{\rm e}^{2\pi {\rm i}\alpha N}}{N}{\rm e}^{{\rm i}\Phi_k}
\label{new_matrix}
\end{equation} 
where a new matrix $N_{kp}$ is
\begin{equation}
 N_{kp}=M_{kp}{\rm e}^{2\pi{\rm i}(k-p+\alpha N)/N}\ .
\end{equation}
Eigenvalues $\Lambda_n^{\prime}$ and eigenvectors $\psi_k(n)$ of the matrix
$N_{kp}$ 
\begin{equation}
 \Lambda_n^{\prime}\psi_k(n)=\sum_{p=0}^{N-1}N_{kp}\psi_p(n)
\end{equation}
can easily be expressed through the eigenvalues and the eigenvectors of the
original matrix $M_{kp}$ 
\begin{equation}
 \psi_k(n)={\rm e}^{2\pi{\rm i}k/N}u_k(n)\;,\;\;\; \Lambda_n^{\prime}={\rm
   e}^{2\pi{\rm i}\alpha}\Lambda_n\;. 
\label{connection}
\end{equation}
But from (\ref{new_matrix}) it follows that matrix $N_{kp}$ is a rank-one
deformation of  matrix $M_{kp}$ so it is possible to construct its eigenvalues
and corresponding eigenvectors in a different way (see e.g. \cite{rank_1}). 

Write a formal expansion of an eigenvector of matrix $N_{kp}$ as a series of
the complete set of eigenvectors of the matrix $M_{kp}$ 
\begin{equation}
 \psi_k(n)\equiv {\rm e}^{2\pi{\rm i}k/N}u_k(n)=\sum_{m=1}^N c_m(n)u_k(m)\ .
\label{psi}
\end{equation}
From (\ref{new_matrix}) one gets
\begin{eqnarray}
\hspace{-2cm}\Lambda_n^{\prime}\sum_{m=1}^N c_m(n)u_k(m)
= \sum_{m=1}^N c_m(n)\Lambda_m u_k(m)-
\frac{1-{\rm e}^{2\pi {\rm i}\alpha N}}{N}{\rm e}^{{\rm i}\Phi_k}\sum_{m=1}^N
c_m(n)\sum_{p=0}^{N-1}u_p(m)\ . 
\end{eqnarray}
Introducing the notations
\begin{equation}
 A_m=\sum_{k=0}^{N-1}u_k(m)\ ,\;\;\;g(n)=\sum_{m=1}^Nc_m(n)A_m
\end{equation}
and using the orthogonality of eigenvectors $u_k(n)$ (\ref{orthogonal}) one
obtains 
\begin{equation}
 c_m(n)=\frac{1-{\rm e}^{2\pi {\rm i}\alpha N}}{N}g(n)
 \frac{B_m}{\Lambda_m-\Lambda^{\prime}_n}\; 
\label{cm}
\end{equation}
where 
\begin{equation}
 B_m=\sum_{k=0}^{N-1}{\rm e}^{{\rm i}\Phi_k}\bar{u}_k(m)\;.
\end{equation}
Multiplying the both sides of (\ref{cm}) by $A_m$ and summing from 1 to $N$
one concludes that every eigenvalues $\Lambda^{\prime}_m$ of the matrix $N_{kp}$
obey the equation 
\begin{equation}
 1=\frac{1-{\rm e}^{2\pi {\rm i}\alpha N}}{N}\sum_{m=1}^N
 \frac{A_mB_m}{\Lambda_m-\Lambda^{\prime}_n}\ . 
\label{equation}
\end{equation}
From (\ref{main}) and (\ref{mu}) it follows that
\begin{equation}
 \Lambda_mB_m^{*}=A_m
\end{equation}
and Eq.~(\ref{equation}) takes the final form
\begin{equation}
 1=\frac{1-{\rm e}^{2\pi {\rm i}\alpha N}}{N}\sum_m
 \frac{\Lambda_m|B_m|^2}{\Lambda_m-\Lambda^{\prime}_n }\ .
\label{final}
\end{equation} 
For rank-one deformations  of a real symmetric matrix all terms in the
corresponding equation would be real and one easily comes to the well known
conclusion that eigenvalues of a rank-one deformation of a real symmetric
matrix are in-between eigenvalues of the unperturbed ones
(cf. \cite{rank_1}). In our case the both matrices, $M_{kp}$ and $N_{kp}$ are
unitary and their eigenvalues lie on the unit circle: $\Lambda_m={\rm e}^{{\rm
    i}\theta_m}$, $\Lambda^{\prime}_n={\rm e}^{{\rm i}\theta^{\prime}_n}$. So
arguments require straightforward modifications. 

\begin{figure}
\begin{center}
\includegraphics[ angle=-90, width=.4\linewidth]{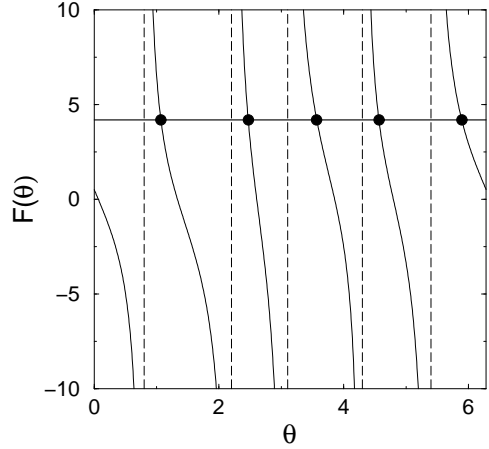}
\end{center}
\caption{Schematic plot of Eq.~(\ref{equation_real}) (solid black
  line). $F(\theta)$ is defined by (\ref{def_F}). Vertical dashed
  lines represent the pole positions. Horizontal straight line indicates the
  value of the right-hand sine of Eq.~(\ref{equation_real}). The abscissa of
  its intersections with $F(\theta)$ (indicated by black circles) give the
  solutions of that equation.}
\label{fig1}
\end{figure}

\begin{figure}
\begin{center}
\includegraphics[width=.4\linewidth]{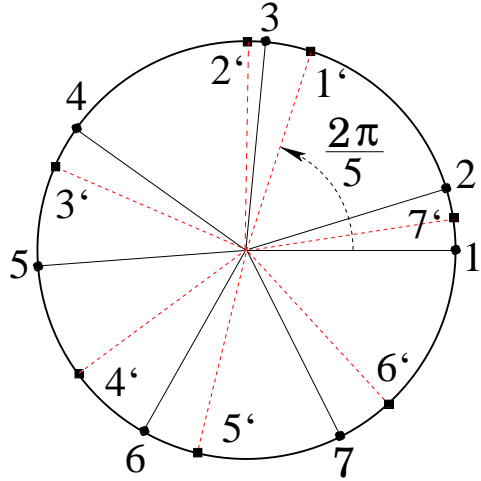}
\end{center}
\caption{Illustration of Lemma 1. Black circles denote the position of 7
  eigenphases for $\alpha=1/5$ and black lines are their radius-vectors. Black
  squares indicate the position of eigenphases after the rotation by angle
  $2\pi/5$ and dashed red lines are radius-vectors of the rotated
  eigenphases. The rotated points are indicated by the same number but with
  sign $'$.  }
\label{fig2}
\end{figure}

From (\ref{final}) one gets
\begin{equation}
 \frac{N}{1-{\rm e}^{2\pi {\rm i}\alpha N}}=\sum_m \frac{|B_m|^2}{1-{\rm
     e}^{{\rm i}(\theta^{\prime}_n-\theta_m)}} 
\end{equation}
which can be rewritten as follows
\begin{equation}
 N(\cot \pi \alpha N-{\rm i})=\sum_m |B_m|^2\left (\cot
   \frac{\theta^{\prime}_n-\theta_m}{2}-{\rm i}\right )\ . 
\label{real_imaginary}
\end{equation}
Due  to the completeness of $u_k(m)$ one has
\begin{equation}
\sum_{n=1}^N\bar{u}_p(n)u_k(n)=\delta_{pk}\;.
\end{equation}
and, consequently,
\begin{equation}
 \sum_{m=1}^N |B_m|^2=N\ .
\label{norm}
\end{equation}
Therefore, the imaginary part of (\ref{real_imaginary}) is identically zero
and new phases $\theta^{\prime}_n$ have to be determined from a real equation 
\begin{equation}
 F(\theta^{\prime}_n)=N\cot \pi\alpha N
\label{equation_real}
\end{equation}
where $F(\theta)$ is defined by:
\begin{equation}
  \label{def_F}
 F(\theta)=\sum_{m=1}^N |B_m|^2\cot \frac{\theta-\theta_m}{2}\;. 
\end{equation}
In the interval $[0,2\pi)$ $F(\theta)$ has poles at
$\theta=\theta_m$ (assuming that all  
$ B_m\neq 0$) and it is monotone between them (cf. Fig.~\ref{fig1}). Let
the eigenphases $\theta_m$ be ordered on the unit circle $0\leq \theta_1\leq
\theta_2\leq \ldots\leq \theta_N<2\pi$. Then between two near-by eigenvalues
$\theta_m$ and $\theta_{m+1}$ (mod $2\pi$) there exists one and only one new
eigenvalue $\theta^{\prime}_n$. Here the new eigenvalues are not necessarily
ordered. 
But according to (\ref{connection}) all eigenphases of the matrix $N_{kp}$
have the form $\theta^{\prime}_m=\theta_m+2\pi \alpha$ (mod $2\pi$). Therefore
we prove the following lemma (cf. Fig.~\ref{fig2}).\linebreak\linebreak
\noindent{\bf Lemma 1.} 
\textit{The eigenvalues of the unitary matrix $M_{kp}$ defined in
  (\ref{matrix}) and (\ref{mu}) are such that after the rotation by
  $2\pi\alpha$ in-between of any pairs of nearest eigenvalues there exist one
  and only one rotated eigenvalue.} 
\linebreak \linebreak 
Multiplying (\ref{new_matrix}) by $\exp( -2\pi{\rm i}(k-p+\alpha N)/N)$ one
gets another  relation
\begin{equation}
  M_{kp}=\tilde{N}_{kp}-\frac{1-{\rm e}^{2\pi {\rm i}\alpha N}}{N}{\rm
    e}^{{\rm i}\Phi_k-2\pi{\rm i}(k-p+\alpha N)/N} 
\label{new_matrix_2}
\end{equation}
with a new matrix $\tilde{N}_{kp}$ 
\begin{equation}
 \tilde{N}_{kp}=M_{kp}{\rm e}^{-2\pi{\rm i}(k-p+\alpha N)/N}
\end{equation}
 whose eigenvalues and eigenvectors are
\begin{equation}
 \tilde{\psi}_k(n)={\rm e}^{-2\pi{\rm i}k/N}u_k(n)\;,\;\;\; 
\tilde{\Lambda}_n^{\prime}={\rm e}^{-2\pi{\rm i}\alpha}\Lambda_n\;.
\end{equation}
Repeating the above calculations but for the matrix $\tilde{N}_{kp}$ one finds
that its eigenvalues $\tilde{\Lambda}^{\prime}_m$ have to be determined from the
equation 
\begin{equation}
 1=\frac{1-{\rm e}^{2\pi {\rm i}\alpha N}}{N} \sum_{m=1}^N
 \frac{\Lambda_m|\tilde{B}_m|^2}{\Lambda_m-\tilde{\Lambda}^{\prime}_n} 
\label{final_2}
\end{equation} 
where
\begin{equation}
 \tilde{B}_m=\sum_{k=0}^{N-1}\bar{u}_k(m){\rm e}^{{\rm i}\Phi_k-2\pi{\rm i}k/N }
\label{bm_tilde}
\end{equation}
As it has exactly the same form as (\ref{final}) and
$\tilde{\Lambda}_n^{\prime}={\rm e}^{-2\pi{\rm i}\alpha}\Lambda_n$ one comes
to the lemma.\linebreak \linebreak   
\noindent{\bf Lemma $1^{\prime}$.}
\textit{The eigenvalues of $M_{kp}$ are such that after the rotation by
  $-2\pi\alpha$  between two nearest eigenvalues of $M_{kp}$ there exists one
  and only one rotated eigenvalue.}   
\linebreak\linebreak
These lemmas prove the existence of long-range correlations between
eigenvalues of the matrix $M_{kp}$ which are merely a consequence of the fact
that rank-one deformations (\ref{new_matrix}) and (\ref{new_matrix_2}) of the
original matrix (\ref{matrix})  have eigenvalues easily expressible through
eigenvalues of the original matrix.

A few other consequences of this property is worth to mention.  As all $N$
solutions of (\ref{final}) have the form $\Lambda^{\prime}_n=\Lambda_n{\rm
  e}^{2\pi {\rm i}\alpha}$ with $n=1,\ldots, N$ the numerators of this
equation can be found explicitly. From~\ref{appendix} it follows that
\begin{equation}
 \frac{1-{\rm e}^{2\pi {\rm i}\alpha N}}{N}\Lambda_m|B_m|^2=
\frac{\displaystyle\prod_{n=1}^N(\Lambda_m-\Lambda_n {\rm e}^{2\pi {\rm
      i}\alpha} )}{\displaystyle\prod_{s\neq m}(\Lambda_m-\Lambda_s )} 
\label{coefficients}
\end{equation}
which can be rewritten in the real form as follows
\begin{equation}
 |B_m|^2\frac{\sin \pi \alpha N}{N\sin \pi \alpha}=\prod_{n\neq m}
\frac{\sin(\frac{1}{2}(\theta_m-\theta_n-2 \pi
  \alpha))}{\sin(\frac{1}{2}(\theta_m-\theta_n))}\ . 
\end{equation}
Similarly, from (\ref{final_2}) one concludes that
\begin{equation}
 \frac{1-{\rm e}^{-2\pi {\rm i}\alpha N}}{N}\Lambda_m|\tilde{B}_m|^2=
\frac{\displaystyle\prod_{n=1}^N(\Lambda_m-\Lambda_n {\rm e}^{-2\pi {\rm i}\alpha} )}{\displaystyle\prod_{s\neq m}(\Lambda_m-\Lambda_s )}\ .
\label{coefficients_tilde}
\end{equation}
So 
\begin{equation}
 |\tilde{B}_m|^2\frac{\sin \pi \alpha N}{N\sin \pi \alpha}=\prod_{n\neq m}
\frac{\sin(\frac{1}{2}(\theta_m-\theta_n+2 \pi \alpha))}{\sin(\frac{1}{2}(\theta_m-\theta_n))}\ .
\end{equation}

\section{Long-range correlations}\label{long_range}

In the preceding Sections it has been  proved that eigenphases of matrix
(\ref{main}) have a special type of long-range correlations. Namely, when one
rotates all eigenvalues of this matrix by $\pm 2\pi \alpha$ and superimposes
the rotated eigenphases with non-rotated ones then in-between two nearest
eigenvalues of the original matrix there will be one and only one rotated
eigenphase.   
In this Section we investigate certain consequences of such correlations in
more details.  

\begin{figure}
\begin{minipage}{.49\linewidth}
\begin{center}
\includegraphics[width=.8\linewidth]{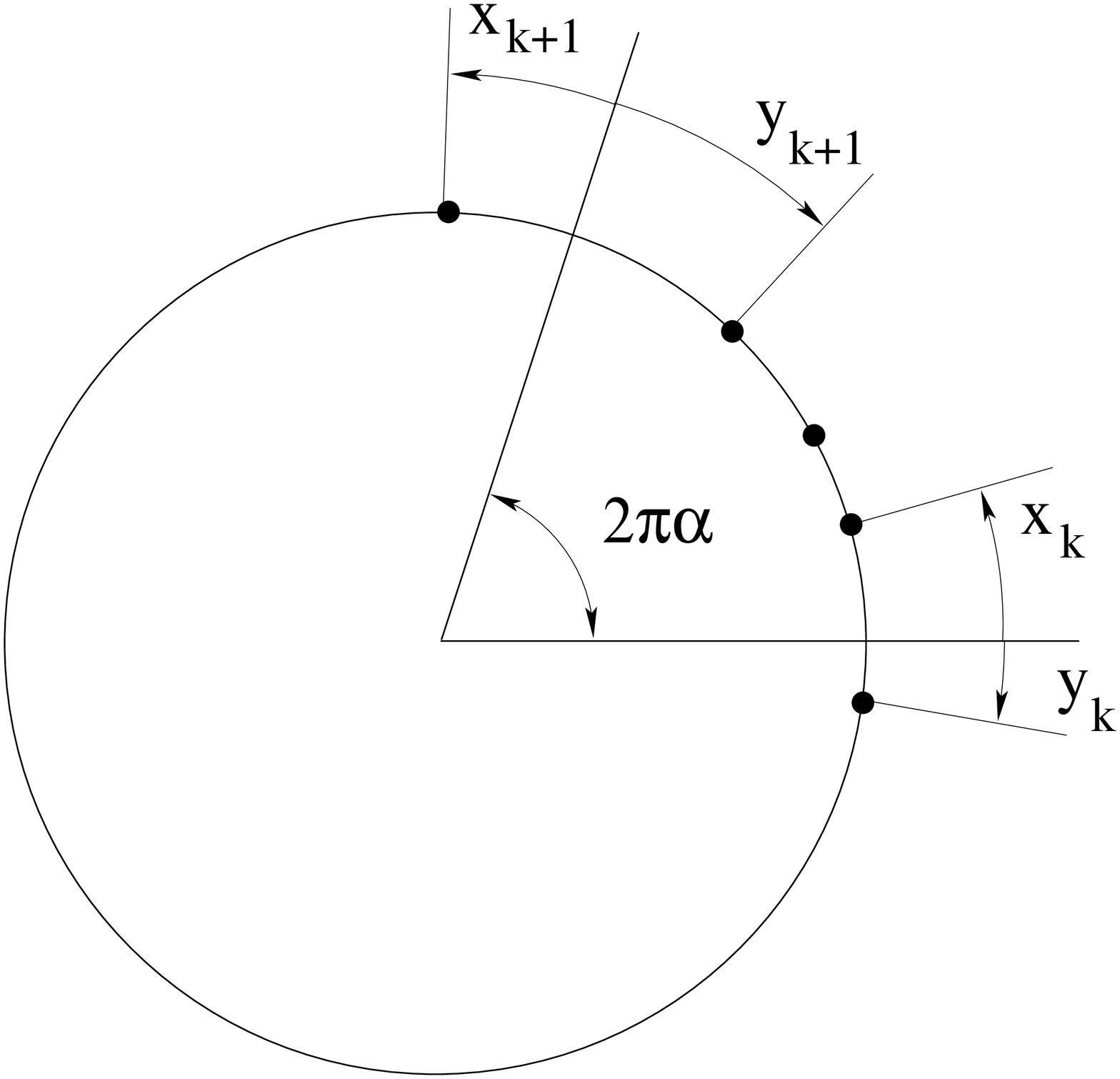}
\end{center}
\begin{center}a)\end{center}
\end{minipage}\hfill
\begin{minipage}{.49\linewidth}
\begin{center}
\includegraphics[width=.84\linewidth]{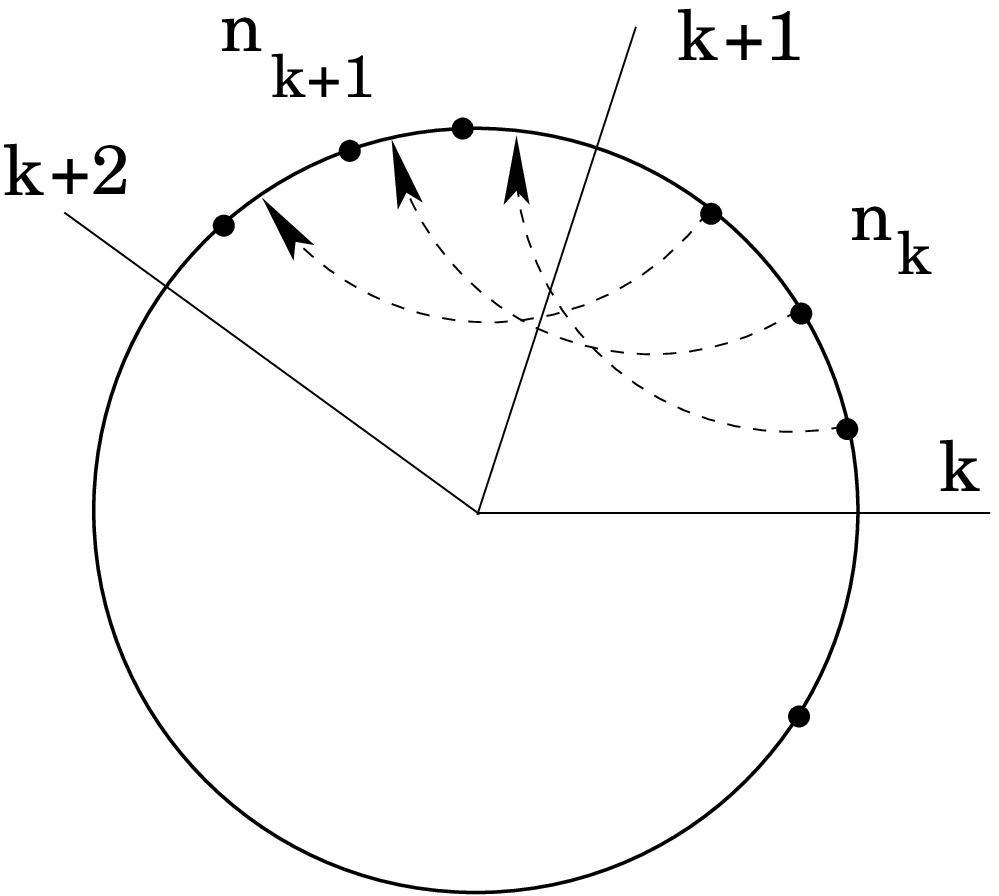}
\end{center}
\begin{center} b) \end{center}
\end{minipage}
\caption{(a) Eigenphases close to the boundaries of a $2\pi \alpha$
  sector. (b) Division of the unit circle into sectors of angle $2\pi
  m/q$. Black circles indicate the position of eigenphases in two near-by
  sectors. Dashed lines show  the positions  which will occupy eigenvalues
  from one sector after the rotation by $2\pi m/q$.} 
\label{recurrence}
\end{figure}

Let us put all eigenvalues of unitary matrix (\ref{matrix}) on the unit circle
and consider a sector of angle $2\pi \alpha$ which contains $n$
eigenvalues (see Fig.~\ref{recurrence}a). 
The sector boundaries divide the unit circle in-between certain
eigenphases. Denote the angular distance from the sector boundaries to the
nearest eigenphases in the clockwise and counterclockwise directions by $y_k$,
$y_{k+1}$ and $x_k$, $x_{k+1}$ respectively
(cf. Fig.~\ref{recurrence}a). After the rotation by $2\pi\alpha$  only 2
possibilities are possible: either  the point $x_k$ or the point $y_k$ will fall inside  the
points $x_{k+1}$ and $y_{k+1}$. In the first
case one has $x_k<x_{k+1}$ and $y_{k+1}<y_k$. In the second case the
inequalities are reversed:  $x_k>x_{k+1}$ and $y_{k+1}>y_k$. Therefore in the
all cases the following inequality is fulfilled 
\begin{equation}
 (y_{k+1}-y_{k})(x_{k+1}-x_{k})<0\ .
\label{inequality}
\end{equation}
This inequality is valid for all $\alpha$ and $N$. From now on we shall
consider only rational $\alpha$ 
\begin{equation}
 \alpha=\frac{m}{q}
\end{equation}
with co-prime  integers $m$ and $q$. 

As above divide the unit circle into $q$ radial sectors of angle $2\pi
m/q$. When $m=1$ these sectors are disjoint but for $m>1$ they will
overlap. Denote the number of eigenphases in the $k^{\mbox{th}}$ sector by
$n_k$ (see Fig.~\ref{recurrence}b).  

After the rotation by $2\pi m/q$ the eigenphases from the $k^{\mbox{th}}$
sector will move into the $(k+1)^{\mbox{th}}$ sector. These rotated points will
divide this sector into $n_k+1$ intervals. According to the lemma $1$
the eigenphases in the $(k+1)^{\mbox{th}}$ sector have to be intertwined with the
rotated eigenvalues. Therefore, all intervals except the first and the last
have to be occupied. The first will be occupied if $x_k>x_{k+1}$ and the last
will be occupied if $y_{k+1}>y_{k+2}$. All these requirements can be rewritten
as the following recurrence relation 
\begin{equation}
n_{k+1}=n_k-1+\Theta(x_{k}-x_{k+1})+\Theta(y_{k+1}-y_{k+2})\ .
\label{rec} 
\end{equation}
Here as in Fig.~\ref{recurrence}a $x_k$ and $y_k$ are distances from the
boundary of the  $k^{\mbox{th}}$ sector to the two closest eigenphases to it
and $\Theta(x)$ is the Heaviside function: $\Theta(x)=1$ if $x>0$ and
$\Theta(x)=0$ if $x<0$. From (\ref{inequality}) it follows that the difference
$y_{k+1}-y_{k+2}$ is of the opposite sign than the the difference
$x_{k+1}-x_{k+2}$ and  the last relation takes the form 
\begin{equation}
n_{k+1}=n_k-1+\Theta(x_{k}-x_{k+1})+\Theta(x_{k+2}-x_{k+1})\  .
\label{recursive} 
\end{equation}
Now choose the beginning of the first sector at the position of an eigenphase
(i.e. impose $y_1=0$). Then from (\ref{inequality}) it follows that $x_2<x_1$ and 
$x_q<x_1$. Direct applications of (\ref{recursive}) give
\begin{equation}
 n_2=n_1-1+\Theta(x_1-x_2)+\Theta(x_3-x_2)=n_1+\Theta(x_3-x_2)
\end{equation}
because $x_1>x_2$,
\begin{eqnarray}
\hspace{-2cm} n_3&=&n_2-1+\Theta(x_2-x_3)+\Theta(x_4-x_3)\nonumber\\
\hspace{-2cm}&=&
 n_1+\Theta(x_3-x_2)-1+\Theta(x_2-x_3)+ \Theta(x_4-x_3)= n_1+\Theta(x_4-x_3) 
\end{eqnarray}
because $\Theta(x)+\Theta(-x)=1$ and so on. In this manner one concludes that
for $j=2,\ldots,q-1$ 
\begin{equation}
n_j=n_1+\Theta(x_{j+1}-x_j) 
\label{nj}
\end{equation}
and $n_q=n_1+1$ because, as was noted above, $x_q<x_1$. 

As the sum over all $q$ sectors cover the unit circle $m$ times, the sum over
all $n_k$ equals $mN$:  $\displaystyle\sum_{k=1}^q n_k=mN$. Therefore  
\begin{equation}
mN=q n_1+1+\sum_{j=2}^{q-1}\Theta(x_{j+1}-x_j)\ .
\label{total_sum}
\end{equation}
When $mN\equiv 1$ mod $q$, all $\Theta$-functions in the right-hand side of
this expression are forced to be zero which leads to the conclusion that in this case
\begin{equation}
 x_q<x_{q-1}<\ldots <x_2<x_1\ .
\label{plus}
\end{equation}
When $mN\equiv -1$ mod $q$, all $\Theta$-functions have to be equal to 1 and
the inequalities are reversed
\begin{equation}
 x_2<x_3<\ldots < x_{q-1}<x_q<x_1\ .
\label{minus}
\end{equation} 
Inequalities (\ref{plus}) and (\ref{minus})  manifest the existence of an
exceptionally strong long-range correlations  when $mN\equiv \pm 1$ mod
$q$. For usual matrix ensembles correlations between eigenvalues separated by
a large distance tend to zero. But in our  case  eigenphases at distances
$2\pi m k/q$ with $k=1,\ldots,(q-1)$ are not independent but restricted by the
above inequalities.  In \cite{Schmit} only these special cases had been
considered.

\section{Geometrical unfolding}\label{geo_unfolding}

\begin{figure}
\begin{minipage}{.43\linewidth}
\begin{center}
\includegraphics[width=.8\linewidth]{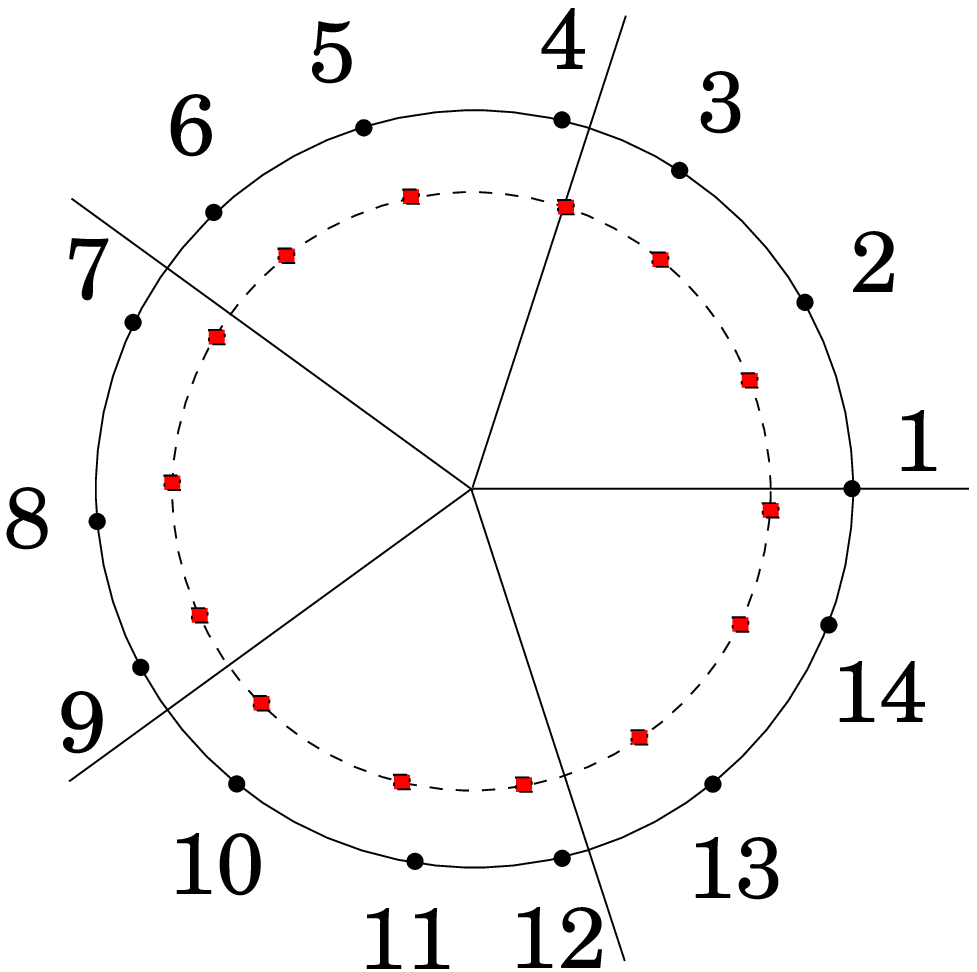}
\end{center}
\begin{center}a)\end{center}
\end{minipage}\hfill
\begin{minipage}{.43\linewidth}
\begin{center}
\includegraphics[width=.89\linewidth]{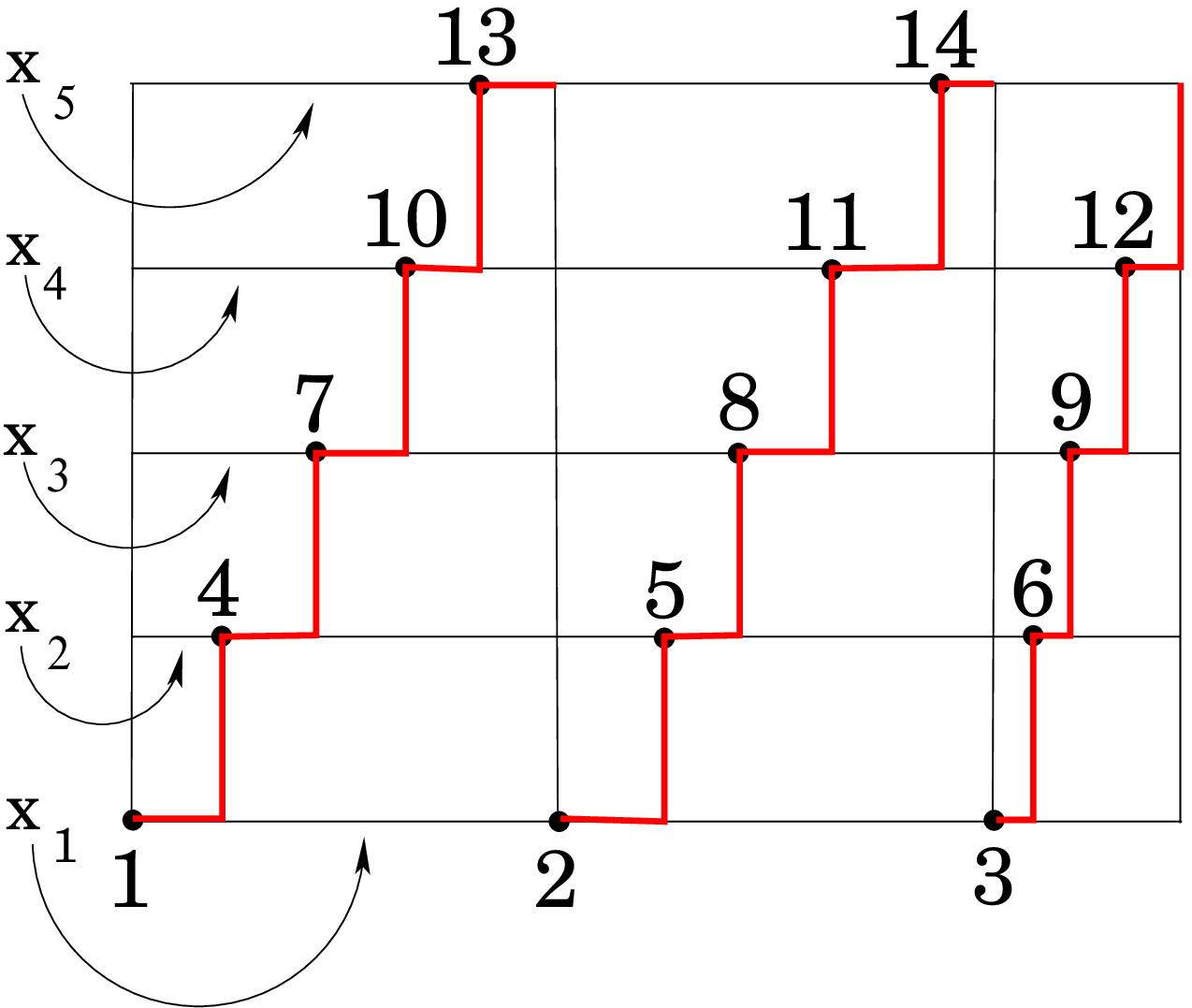}
\end{center}
\begin{center}b)\end{center}
\end{minipage}

\vspace{.5cm}

\caption{(a) Small black circles:  schematic view of eigenvalues of matrix
  (\ref{matrix}) for $\alpha=1/5$ and $N=14$ ($N\equiv -1$ mod $5$).  Numbers
  from 1 to 14 indicate the consecutive eigenphases. Eigenphases rotated by
  $2\pi/5$ are denoted by red squares. For clarity they are situated on a
  smaller dashed circle.  (b) The same configuration of eigenvalues but
  unfolded on 5 horizontal lines representing 5 sectors. $x_1,\ldots,x_5$ are
  the distances from the beginning of the sectors to the closest eigenphase as
  in Fig.~\ref{recurrence}a. Thick red lines demonstrate  relative eigenvalue
  positions.} 
\label{unfolding}
\end{figure}

To visualize better the restrictions implied by inequalities (\ref{inequality})
and  recurrence relations (\ref{recursive}) let us split the unit circle into
$q$ angular sectors of angle $2\pi m/q$ as above and  denote the positions of
eigenvalues inside each sector on $q$ parallel lines (see
Fig.~\ref{unfolding}).  

According to Lemmas 1 and $1^{\prime}$, between two closest eigenphases of the
matrix (\ref{matrix}) there is one and only one eigenphase  rotated by
$2\pi\alpha=2\pi m/q$. In the unfolded description it is manifested by the
condition that points at each line have to be in-between two close-by points
on the lower line. The relative positions of eigenvalues strongly depend on
distances $x_k$ from the beginning of each sector to the eigenphase closest to
it (see (\ref{total_sum})). In the unfolded representation (as in
Fig.~\ref{unfolding}a) $x_k$ are the distances along the horizontal lines to points
closest to a vertical line which represents the boundary of a sector with
angle $2\pi m/q$. 

Let start from the lower left point and draw a horizontal line till there is at
the vertical a point situated at the first line above it. Then draw the
vertical line till it touches that point. Now continue drawing an horizontal
line 
till there is at the vertical a point at the upper line closest to the
boundary of the given sector and so on. This line will go right if
$x_{k+1}>x_k$ and left if $x_{k+1}<x_k$. Finally, points will be connected by
step-wise lines as in Fig.~\ref{unfolding}b. Notice that according to our
convention point 1 does not belong to the first line but to the last one. The
shape of these lines  
are determined by the inequalities between all $x_k$. In Fig.~\ref{unfolding}
the case of $N\equiv -1$ mod $5$ is indicated. According to (\ref{minus})
$x_{k+1}>x_k$ for $k=1,\ldots,4$ which explains the staircase  form of these
lines. They all start at points along the  first  horizontal line, go up and
to the right,  and finally finish at the last horizontal line but with the
shift by 1 unit. It is clear that such lines cannot cross each other.  

\begin{figure}
\begin{minipage}{.43\linewidth}
\begin{center}
\includegraphics[width=.89\linewidth]{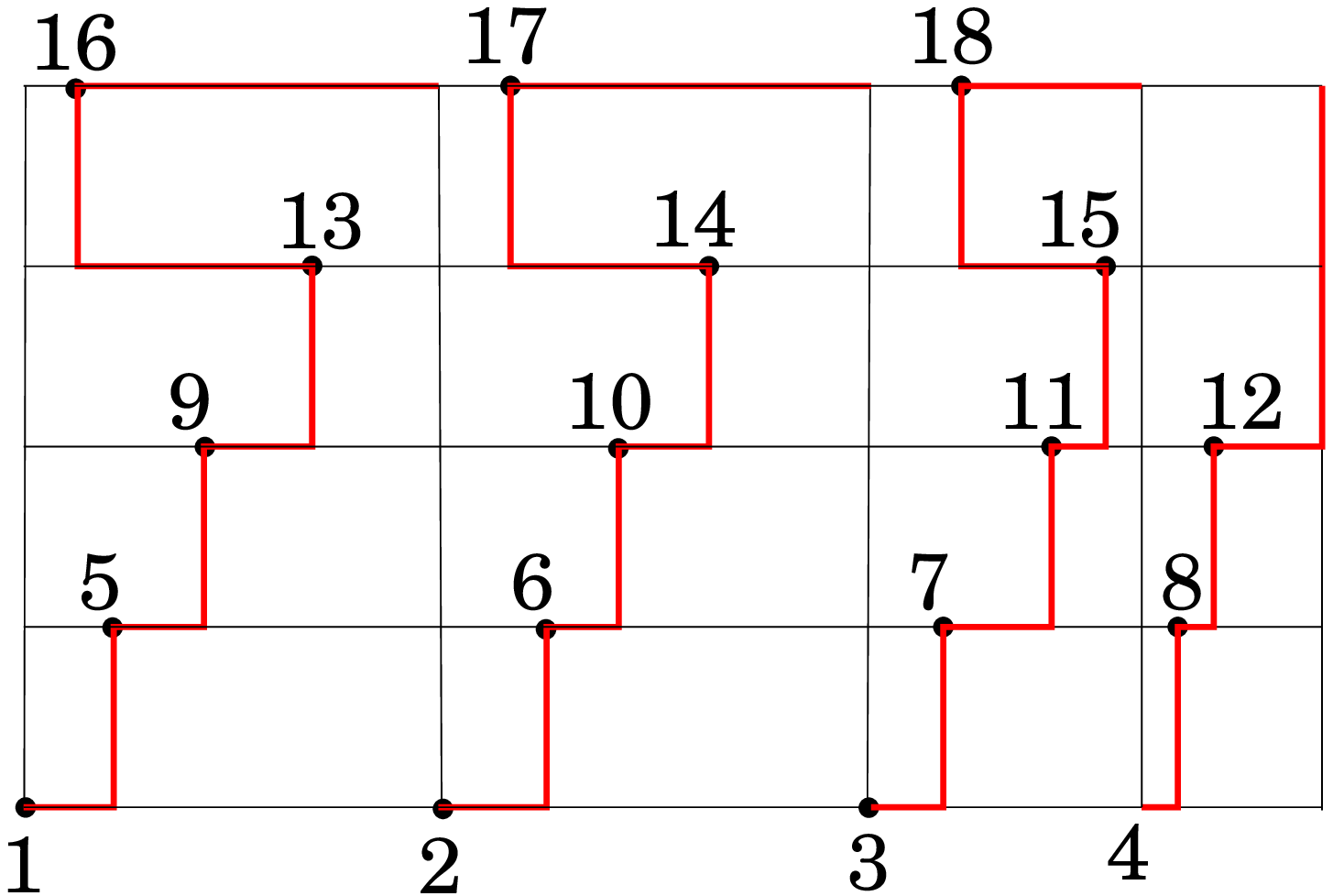}
\end{center}
\begin{center}a)\end{center}
\end{minipage}\hfill
\begin{minipage}{.43\linewidth}
\begin{center}
\includegraphics[width=.89\linewidth]{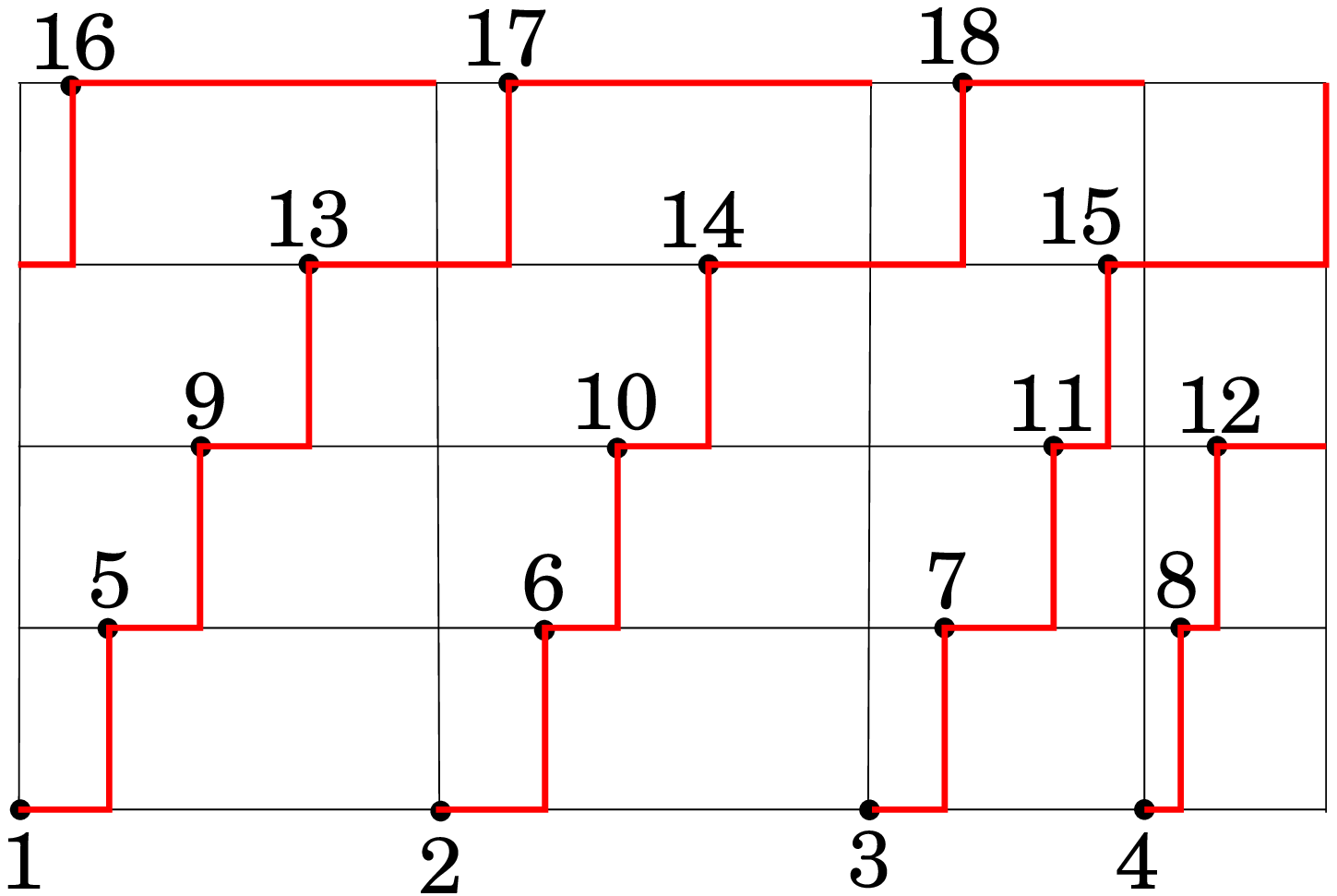}
\end{center}
\begin{center}b)\end{center}
\end{minipage}

\vspace{.5cm}

\caption{(a) Small black circles:  schematic view of eigenvalues of matrix
  (\ref{matrix}) for $\alpha=1/5$ and $N=18$  (i.e. $N\equiv -2$ mod
  $5$). Thick red lines connect eigenphases as indicated in the text.  (b) The
  same configuration of eigenvalues but connected by non-decreasing staircase
  lines.}
\label{unfold_18}

\end{figure}

For other matrix dimensions these lines will have a different shape. Consider
first as an example the case $\alpha=1/5$ and  $N\equiv 3$ mod $5$. From
(\ref{total_sum}) it follows that in this case only 2 of 3 $\Theta$-functions
have to be equal to 1. Every time one of the $\Theta$-function is zero,
the above lines turns to the left (see Fig.~\ref{unfold_18}) so the shape of
the step-wise curve differs from the one of Fig.~\ref{unfolding}b. Consider
the horizontal line when it first turns left. Instead of the left turn let us
continue to the right till we touch an other point on this line.  As between
these two points there exists only one point at the lower and higher lines
there is no contradiction with our line construction
(cf. Fig.~\ref{unfold_18}). Finally we come to the conclusion that for these
values of $\alpha$ and $N$ the eigenphases have to be connected by
non-intersecting lines which  
go only up and to the right and whose initial and final points are shifted by
2 units.  

These arguments can be generalized  for all values of matrix dimensions and we
get the following lemma.\\ \linebreak 
\noindent{\bf Lemma 3}: \textit{For $\alpha=m/q$ and $mN\equiv k$ mod $q$ with
  $k=1,\ldots,q-1$  mutual positions of eigenphases of matrix (\ref{matrix})
  can be described as follows. 
Fix $q$ horizontal lines, put arbitrary  points  at the lowest line, and
notice the vertical images of these points along the last line.  Draw
staircase non-intersecting lines going  only up and to the right with the
condition that they start at the lower line and end at last line but with the
shift by $k$ units.  Points at horizontal lines are situated at the corners of
the constructed lines.} 

When $k>q/2$ one may simplify the construction by using  non-intersecting
step-wise lines going up and to the left with the shift by $q-k$.  It
implies that properties of the cases $mN\equiv k$ mod $q$ and $mN\equiv -k$
mod $q$ are the same.  

\section{Transfer operator}\label{transfer}

The above unfolding gives not only  a clear picture of mutual positions of
eigenphases  but also serves as  the basis of the explicit calculation of the
spectral 
statistics for the problem under consideration. The calculations are based on
the following conjecture  proposed in \cite{Schmit}. 

\textbf{Conjecture}: \textit{For $\alpha=m/q$ the eigenvalues of the
  $q^{\mathrm{th}}$ power of the original matrix (\ref{matrix}) for all $N\not
  \equiv  0$ mod $q$ have universal  spectral statistics independent on $q$
  and $N$ but different  for different symmetry classes of random phases
  $\Phi_k$. For non-symmetric ensemble the statistics of $M^q$ coincides with
  the Poisson statistics and for symmetric ensemble (\ref{symmetry}) it is
  described by the semi-Poisson statistics with $\beta=-1/2$ called in
  \cite{Schmit} the super-Poisson statistics.}

The main physical argument in favor of this conjecture in \cite{Schmit} was
the fact that for rational $\alpha=m/q$ the $q^{\mathrm{th}}$ power of the
classical map (\ref{map}) corresponds to a classically integrable map, and
according to the usual wisdom  \cite{Berry} integrable models have to be
described by the Poisson statistics. Extensive numerical calculations agree
very well with this conjecture. But it seems that  to prove it rigorously one
has to develop new methods which are at present under investigation
\cite{BogomolnyGiraud}.  

After unfolding, when the points from all the sectors separated by $2\pi m/q$
are taken together, they can be considered as the eigenphases of the
$q^{\mathrm{th}}$ power of the original matrix (with evident
rescaling). Assuming the validity of the conjecture it follows that for all
$N\not \equiv 0$ mod $q$ these points constitute the semi-Poisson ensemble
with $\beta=0$ for non-symmetric matrices and $\beta=-1/2$ for symmetric ones.   

In particular, the probability that between two eigenvalues of $M^{q}$
separated by $x$ there exist exactly $r$  eigenvalues is the following
\begin{equation} 
\hspace{-2cm} p_r(x)=\left \{\begin{array}{cl}\displaystyle\frac{x^r}{r!}\mathrm{e}^{-x}
    &\mathrm{for\; non-symmetric\; ensemble}\\  
\displaystyle\frac{1}{2^{(r+1)/2}\Gamma((r+1)/2)}x^{(r-1)/2}\mathrm{e}^{-x/2}
&\mathrm{for\; symmetric\; ensemble} \end{array}\right . .
\label{probability_r}
\end{equation}
The results of the preceding Section can be reformulated such that the joint
probability of the close-by levels integrated over all the possible positions of
other
levels is the same as the probability of non-intersecting staircase lines which
start from the initial levels and which finish  after $q$ steps (where $q$
is the denominator of $\alpha$) with the shift of $k$ units (where $k$ is the
residue of $mN$ modulo $q$). According to the conjecture the distribution of
unfolded points are known which permits the  calculation of spectral
statistics of the original matrix (\ref{matrix}).  

The usual method of computing the probability of non-intersecting paths in a
Markoff process is the determinantal representation \cite{determinant}. We
found that for practical reasons it is more convenient to use the transfer
operator method. In this method one first unfolds the spectrum as indicated in
Fig.~\ref{unfold_18}. Then one considers a vertical strip bounded by vertical
lines emanating from any two near-by levels of original matrix, say $x_2$ and
$x_3$ in Fig.~\ref{unfold_18} separated by the distance $x= x_3-x_2$. Now
different horizontal lines  can enter and can leave the strip. When all
in-coming and out-coming horizontal lines are fixed,  it is obvious that the
configuration inside the strip is not affected by outside points. Therefore,
it is possible to integrate over all configurations of points inside the strip
with prescribed ordering.  Denoting the initial and final  lines by indexes
$i$ and $j$ the result of integration constitutes the matrix element
$T_{ji}(x)$ of the transfer matrix $T(x)$.  

For rational $\alpha=m/q$ and $mN\equiv k$ mod $q$ with $1\leq k\leq q-1$ each
initial (and  final) state is determined by fixing $l=q-k-1$ horizontal lines
\begin{equation}
1\leq i_1<i_2<\ldots <i_{l}\leq q-2
\label{lines}
\end{equation}
from the total number of lines equal $q-2$.  
It means that the dimension of the transfer operator is
\begin{equation}
t=C_{q-2}^{l}
\end{equation}
and it is convenient to label the set of $l$ integers obeying (\ref{lines}) in
e.g. lexicographical order.  

\begin{figure}
\begin{minipage}{.3\linewidth}
\begin{center}
\includegraphics[width=.8\linewidth]{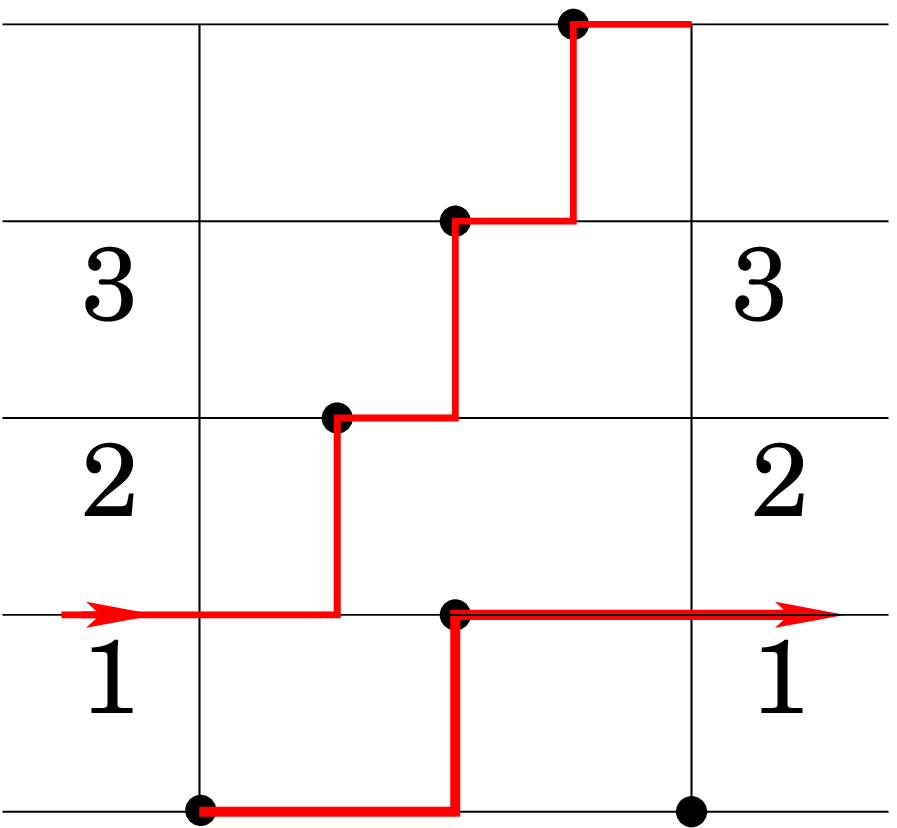}
\end{center}
\begin{center}$T_{11}$\end{center}
\end{minipage}\hfill
\begin{minipage}{.3\linewidth}
\begin{center}
\includegraphics[width=.8\linewidth]{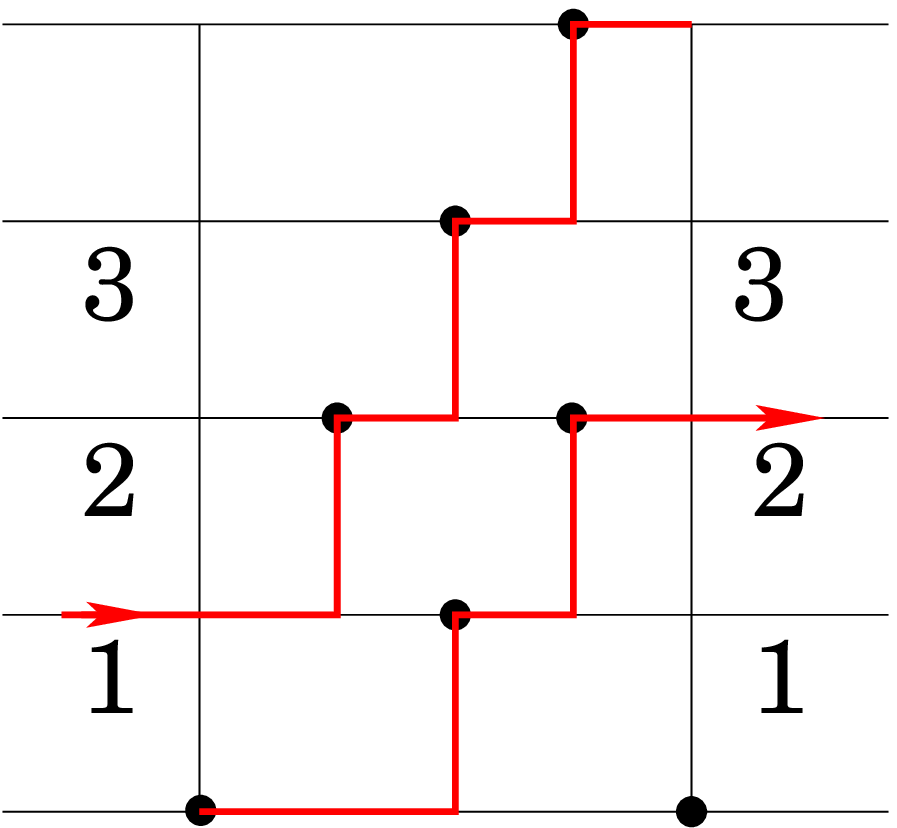}
\end{center}
\begin{center}$T_{12}$\end{center}
\end{minipage}\hfill
\begin{minipage}{.3\linewidth}
\begin{center}
\includegraphics[width=.8\linewidth]{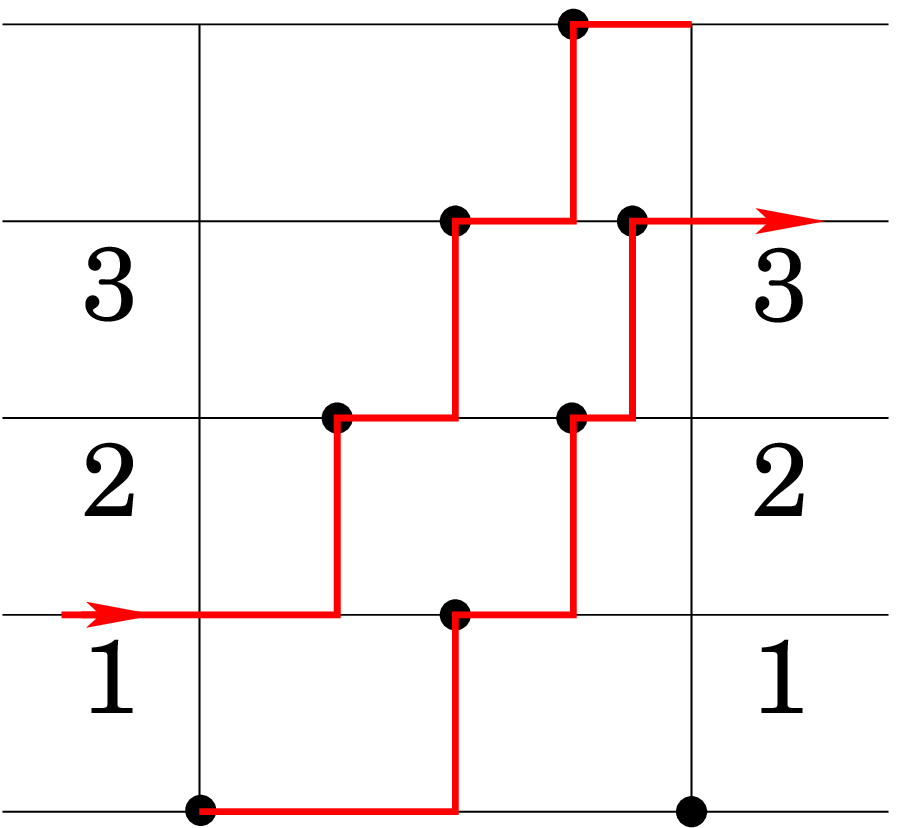}
\end{center}
\begin{center}$T_{13}$\end{center}
\end{minipage}

\vspace{.5cm}

\begin{minipage}{.3\linewidth}
\begin{center}
\includegraphics[ width=.8\linewidth]{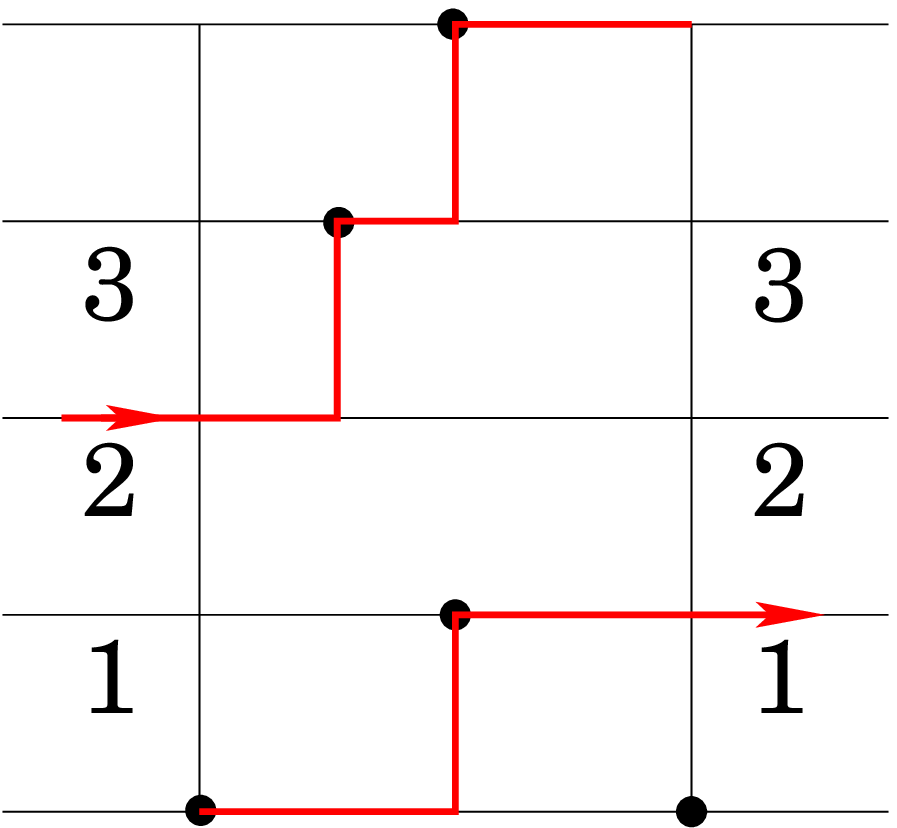}
\end{center}
\begin{center}$T_{21}$\end{center}
\end{minipage}\hfill
\begin{minipage}{.3\linewidth}
\begin{center}
\includegraphics[width=.8\linewidth]{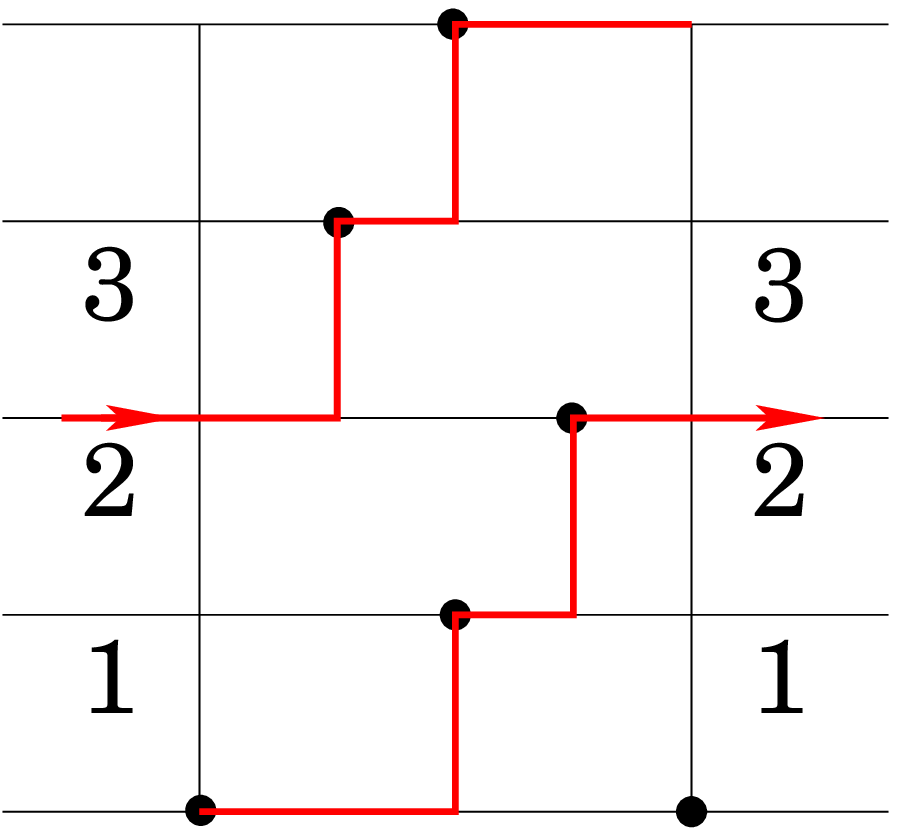}
\end{center}
\begin{center}$T_{22}$\end{center}
\end{minipage}\hfill
\begin{minipage}{.3\linewidth}
\begin{center}
\includegraphics[width=.8\linewidth]{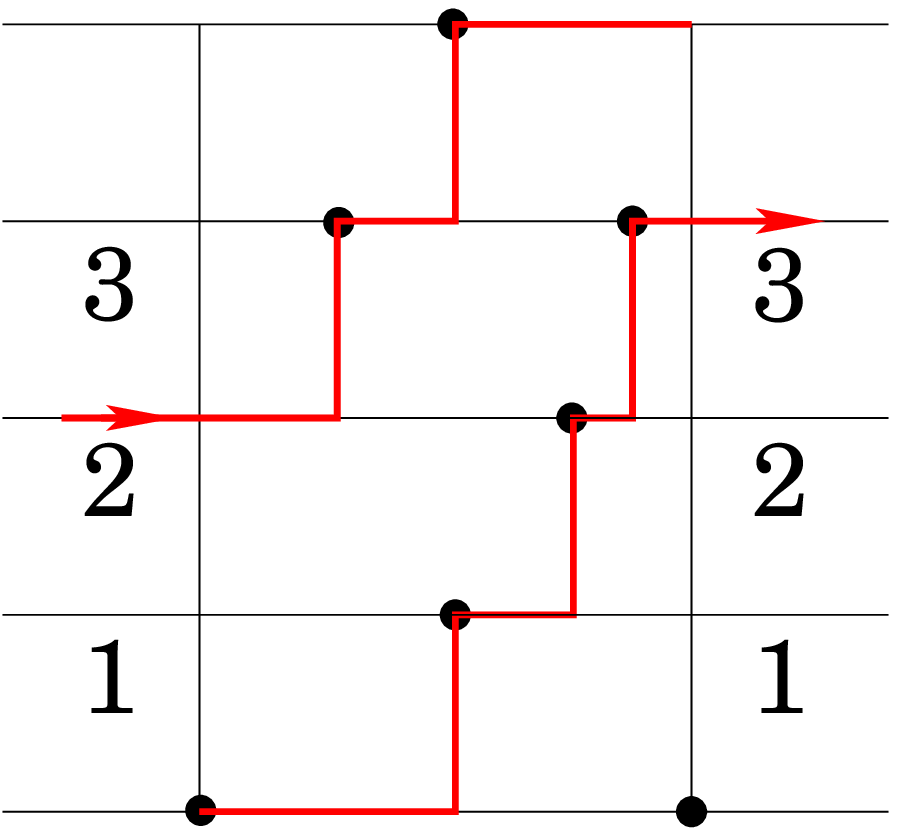}
\end{center}
\begin{center}$T_{23}$\end{center}
\end{minipage}

\vspace{.5cm}

\begin{minipage}{.3\linewidth}
\begin{center}
\includegraphics[width=.8\linewidth]{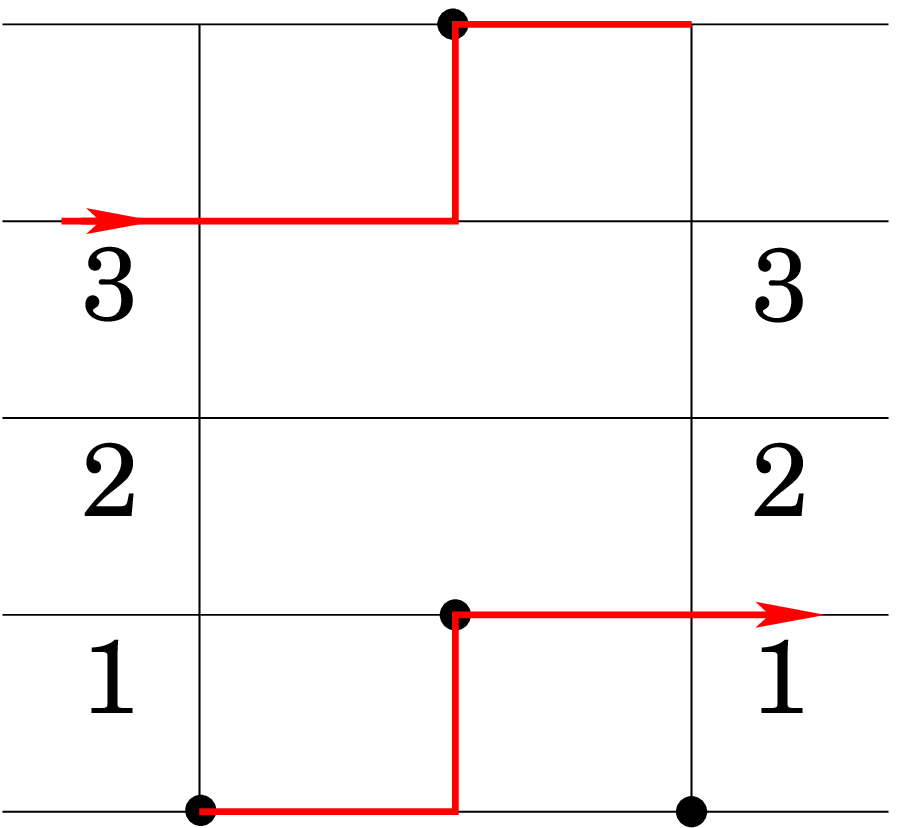}
\end{center}
\begin{center}$T_{31}$\end{center}
\end{minipage}\hfill
\begin{minipage}{.3\linewidth}
\begin{center}
\includegraphics[width=.8\linewidth]{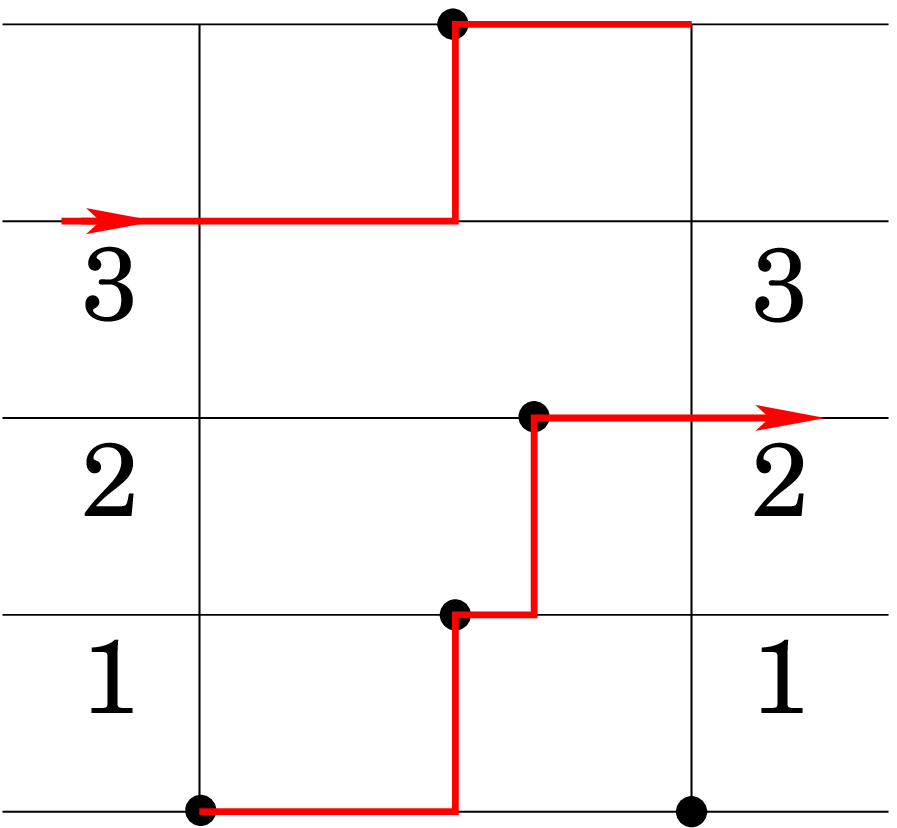}
\end{center}
\begin{center}$T_{32}$\end{center}
\end{minipage}\hfill
\begin{minipage}{.3\linewidth}
\begin{center}
\includegraphics[width=.8\linewidth]{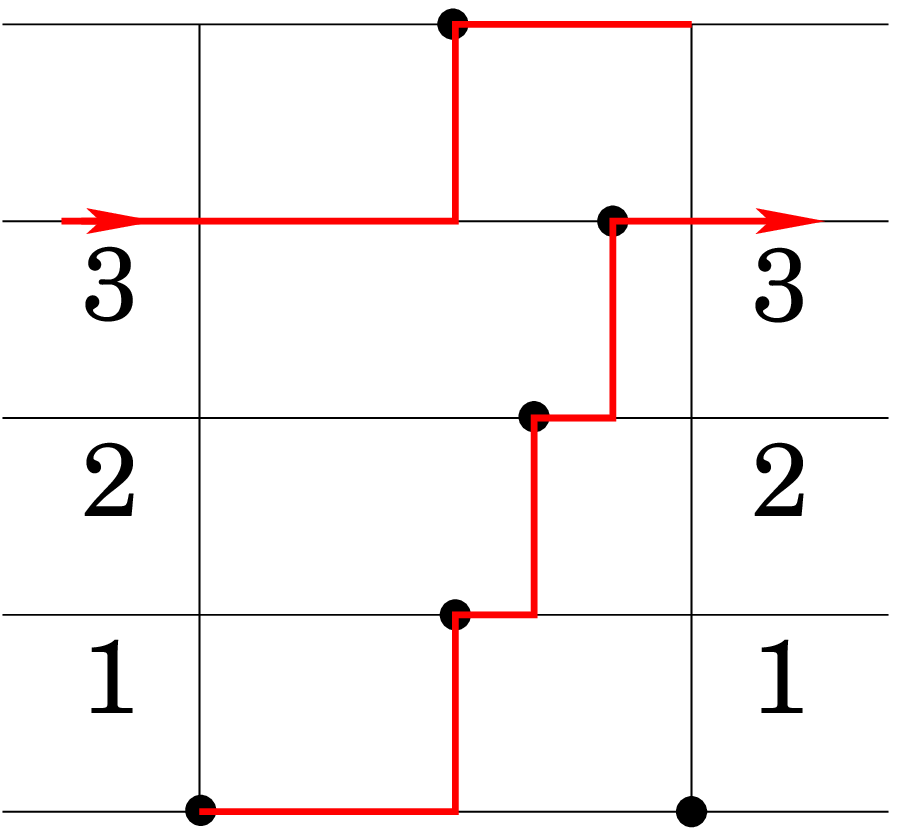}
\end{center}
\begin{center}$T_{33}$\end{center}
\end{minipage}

\vspace{.5cm}

\caption{Structure of the transfer operator for $\alpha=1/5$ and $N\equiv -2$
  mod $5$.} 
\label{m_ij}
\end{figure}

For clarity, let us consider the case $\alpha=1/5$ and $N\equiv 3$ mod $5$ in
detail ($l=1$). As $q=5$ there is one possible horizontal line which may go
through 
the vertical strip (cf. Fig.~\ref{unfold_18}). Therefore the transfer matrix
is $3\times 3$ matrix labeled by the number of these lines. In Fig.~\ref{m_ij}
all possible configurations for this case are presented.   

In general, if $i\equiv (i_1,i_2,\ldots,i_{l})$ and $j\equiv
(j_1,j_2,\ldots,j_{l})$ are multi-indexes of an initial and a final states,
the total number of points, $r$, inside the considered vertical strip is
determined by the expression 
\begin{equation}
r=j_1+\sum_{s=1}^{l-1}[j_{s+1}-i_{s}]+q-1-i_{l}=|j|-|i|+q-1
\label{rtot}
\end{equation}  
where $|j|\equiv j_1+\ldots+j_{l}$ and  $|i|\equiv i_1+\ldots+i_{l}$. In general, $r_{\mathrm{min}}\leq r\leq r_{\mathrm{max}}$  with 
\begin{equation}
r_{\mathrm{min}}=2\ ,\;\; r_{\mathrm{max}}=k(q-k)\ .
\label{limits}
\end{equation}
If at least one term in the square brackets in (\ref{rtot}) is negative, the
configuration is impossible and the corresponding matrix element equals
zero. Otherwise, the integration over all intermediate configurations
compatible with the imposed inequalities gives the value of the transfer
matrix elements.  

The calculation of this probability is straightforward. According to the above
conjecture, the probability that between 2 eigenvalues of $M^q$ separated by
$x$ there exist $r$ ordered eigenvalues $y_s$ such that 
\begin{equation}
0\leq y_1\leq y_2\leq \ldots \leq y_r\leq x
\end{equation}  
 is given by (\ref{probability_r}). Therefore, the transfer matrix element
 $T_{ji}(x)$ is the product of 2 factors 
\begin{equation}
T_{ji}(x)=n_{ji}p_r(x)
\end{equation}
where $r$ is the integer determined by (\ref{rtot}), $p_r(x)$ is the same as
in (\ref{probability_r}), and $n_{ji}$ is the number of configurations of $r$
points which fulfilled all inequalities comparable with the fixed initial and
final states. Interchanging the initial and final states and counting
horizontal lines from the top one gets that  the transfer   operator matrix
elements obey the following symmetry 
\begin{equation}
T_{ji}(x)=T_{i^T j^T}(x)
\label{symT}
\end{equation}  
where if $i=(i_1,\ldots,i_l)$, $i^T=(q-1-i_l,\ldots,q-1-i_1)$.

For example,  for $\alpha=1/5$ and $N\equiv -2$ mod $5$ the $T_{12}$ element
includes 5 points (cf. Fig.~\ref{m_ij}), 3 points, $a$, $b$, $c$ belong to the
upper curve, and 2 points, $A$ and $B$,  belong to the lower curve. From the
mutual positions of these points it follows that the $T_{12}(x)$ matrix
element equals the probability that the following inequalities are fulfilled 
\begin{equation} 
0\leq a \leq b \leq c \leq x ,\;\;0\leq A\leq B \leq x ,\;\;a \leq A ,\;\; b
\leq B . 
\label{inequalities}
\end{equation}
By inspection one finds that  there exist exactly 5 possible ordered sequences
compatible with inequalities (\ref{inequalities}), namely 
$$
a\ b\ c\ A\ B,\;\;\;
a\ b\ A\ B\ c,\;\;\;
a\ A\ b\ B\ c,\;\;\;
a\ b\ A\ c\ B,\;\;\;
a\ A\ b\ c\ B .\;\;\;
$$ 
Therefore $T_{12}(x)=5p_5(x)$. Following the symmetry (\ref{symT}) we also
have $T_{23}(x)=~T_{12}(x)=~5p_5(x)$. 

By construction the joint probability of the eigenvalues is equal to the product
of the transfer matrices over all near-by points. As we are interested in the
limit of large number of eigenvalues, the exact behavior near the boundaries
are not important and one can take simply the trace of the whole
product. Finally the joint probability of near-by levels of the original
matrix $0<x_1<x_2<\ldots <x_K<L$ integrated over all possible configuration of
levels on other sectors takes the form  
\begin{equation}
\hspace{-2cm}P_L(x_1,\ldots, x_K)\sim \mathrm{Tr}\left [T(x_{K}-x_{K-1})\cdot
  \ldots \cdot 
  T(x_2-x_1) )\right ]\delta(x_1+x_2+\ldots +x_K-L)\ . 
\label{Laplace}
\end{equation} 
The further steps are standard (see e.g. \cite{Gerland}). One has first to
calculate the Laplace transform of the transfer matrix 
\begin{equation}
\hat{T}(k)\equiv \int_0^{\infty}T(x)\mathrm{e}^{-kx}\mathrm{d}\,x\ .
\end{equation} 
Let $\lambda(k)$ be the largest eigenvalue of  $\hat{T}(k)$,  and $w(k)$,
$v(k)$ be respectively the right and left eigenvectors of  this matrix
corresponding to $\lambda(k)$ 
\begin{equation}
\hat{T}(k)w(k)=\lambda(k) w(k)\;,\;\;^{t}v(k)^{t}\hat{T}(k)=\lambda(k) ^{t}v(k)\  .
\label{hat_t}
\end{equation}
In the limit $K\to\infty$ the dominant contribution comes \cite{Gerland} from
a vicinity of the saddle point, $k_{\mathrm{sp}}=h$, defined from the
condition of the fixed mean level density $L/K$ which we normalize to 1  
\begin{equation}
\frac{\lambda^{\prime}(h)}{\lambda(h)}+1=0\ . 
\label{saddle}
\end{equation}
Then the nearest-neighbour distribution is determined by the formula
\begin{equation}
p(s)=\frac{\mathrm{e}^{-hs}}{\lambda(h)}\frac{^{t}v(h)T(s)w(h)}{^{t}w(h)v(h)}\ .
\label{nearest}
\end{equation}
Condition (\ref{saddle})  is equivalent to  the standard normalization of  $p(s)$ 
\begin{equation}
\int_0^{\infty}p(s)\mathrm{d}s=1,\;\;\int_0^{\infty}sp(s)\mathrm{d}s=1.
\end{equation}
From (\ref{probability_r}) it follows that 
\begin{equation}
\hat{T}_{ji}(k)=n_{ji}\left  \{\begin{array}{ll}(k+1)^{-(r+1)},&\mathrm{for\;
      non-symmetric\; ensemble }\\ 
(2k+1)^{-(r+1)/2},&\mathrm{for\; symmetric\; ensemble}\end{array}\right . .
\end{equation}
As $r$ is determined by (\ref{rtot}), the dependence of eigenvalues and eigenvectors (\ref{hat_t}) on $k$ is easy to find
\begin{equation}
w_{i}=\tilde{w}_i (k+1)^{-|i|}\ ,\;\;v_i=\tilde{v}_i (k+1)^{|i|}\ ,\;\;\lambda(k)=\tilde{\lambda}(k+1)^{-q}
\end{equation}
for non-symmetric matrices, and 
\begin{equation}
w_{i}(k)=\tilde{w}_i (2k+1)^{-|i|/2}\ ,\;\;v_i(k)=\tilde{v}_i (2k+1)^{|i|/2}\ ,\;\;\lambda(k)=\tilde{\lambda}(2k+1)^{-q/2}
\end{equation}
for symmetric matrices. Here $i$ denotes the multi-index $(i_1,\ldots
,i_{l})$, $|i|=i_1+\ldots +i_{l}$, and all tilded quantities are
independent on $k$. 

From these relations one finds that the saddle point $h$ obeying (\ref{saddle}) is
\begin{equation}
h=\left  \{\begin{array}{ll}q-1&\mathrm{for\; non-symmetric\; ensemble }\\
(q-1)/2&\mathrm{for\; symmetric\; ensemble}\end{array}\right . .
\end{equation}
Using (\ref{limits}) we conclude that the nearest-neighbour distribution
(\ref{nearest})  for $\alpha=m/q$ and $mN\equiv k$ mod $q$   equals the
following finite sums 
\begin{equation}
p(s)= \sum_{n=2}^{k(q-k)}a_n s^n \mathrm{e}^{-qs}
\end{equation}
for non-symmetric matrices and
\begin{equation}
p(s)= \sum_{n=1}^{(k(q-k)-1)}a_{n/2} s^{n/2} \mathrm{e}^{-qs/2}
\end{equation}
for symmetric ones. 

The nearest-neighbour distribution for all considered cases (with $k\neq 0,\pm
1$ mod $q$) manifests level repulsion at small $s$ 
\begin{equation}
p(s)\sim \left  \{\begin{array}{ll}s^2&\mathrm{for\; non-symmetric\; ensemble }\\
s^{1/2}&\mathrm{for\; symmetric\; ensemble}\end{array}\right . 
\end{equation}
and has the exponential decrease at large $s$ as it should be for intermediate statistics.

Other correlation functions can also be written explicitly through the same
quantities \cite{Gerland}. In particular, the two-point correlation form factor
has the following form 
\begin{equation}
K(\tau)=1+2\mathrm{Re}\, g(2\pi{\rm i}\tau)
\end{equation}
where 
\begin{equation}
g(t)=\displaystyle\frac{^{t}w(h)L(t+h)(1-L(t+h))^{-1}v(h)}{^{t}w(h)v(h)}
\end{equation}
and the matrix $L(s)=\hat{T}(s)/\lambda(h)$. 

One can check that for all $N$ the level compressibility $K(0)=1/q$ for non-symmetric matrices and $K(0)=2/q$ for symmetric ones.  

Numerically it was established \cite{georgeot} but not yet proved
analytically  that eigenvectors of the considered ensembles of random
matrices have fractal properties independent on the residue $k\neq 0$ mod $q$.

\section{Explicit calculations}\label{explicit}

The simplest new case corresponds to $\alpha =1/5$ and $N\equiv \pm 2$ mod
$5$. Considering  all configurations in Fig.~\ref{m_ij}, one gets that  in
this case  the transfer matrix has the following form  
\begin{equation}
T(x)=\left ( \begin{array}{ccc}3p_4(x)&5p_5(x)&5p_6(x)\\
                            3p_3(x) &5p_4(x)&5p_5(x)\\
                            2p_2(x)&3p_3(x)&3p_4(x)
              \end{array}\right )\ .               
\end{equation}
Performing the calculations discussed in the precedent Section we find that
when $\alpha=1/5$ and $N\equiv \pm 2$ mod $5$ the nearest-neighbour
distribution for non-symmetric matrices is 
\begin{equation}
p(s)=(a_2s^2+a_3s^3+a_4s^4+a_5s^5+a_6s^6)e^{-5s}
\label{ps_5}
\end{equation}
where coefficients $a_n$ are the following: 
$a_2=625/2-275\sqrt{5}/2\approx 5.041,$
$a_3=3125/2-1375\sqrt{5}/2\approx 25.203,$
$a_4=71875/48+33125\sqrt{5}/48\approx 45.724,$
$a_5=-15625/3+9375\sqrt{5}/4\approx 32.451,$
$a_6=1015625/288-453125\sqrt{5}/288\approx 8.357.$

In a similar manner one finds  that for symmetric matrices under the same conditions $p(s)$ is given by the following expression 
\begin{equation}    
p(s)=(a_{1/2}s^{1/2}+a_1s+a_{3/2}s^{3/2}+a_2s^2+a_{5/2}s^{5/2})e^{-5s}
\label{pss_5}
\end{equation}
with $a_{1/2}\approx .3597,\;a_{1}\approx 1.5122,\;a_{3/2}\approx 2.6105,\;a_{2}\approx 1.9471,\;a_{5/2}\approx .5725$. 

For $\alpha=1/7$ and $N\equiv \pm 2$ mod $7$ the transfer operator is represented by  the  $5\times 5$ matrix:
\begin{equation}
T(x)=\left(
\begin{array}{ccccc}
  5 p_6(x) & 14 p_7(x) &28 p_8(x) & 42 p_9(x)&  42 p_{10}(x) \\
  5 p_5(x) & 14 p_6(x) & 28 p_7(x) & 42 p_8(x) & 42 p_9(x) \\
  4 p_4(x)&  10 p_5(x) & 19 p_6(x) & 28 p_7(x)&  28 p_8(x) \\
  3 p_3(x)  & 6 p_4(x) & 10 p_5(x) & 14 p_6(x) & 14 p_7(x)\\
  2 p_2(x)  & 3 p_3(x) &  4 p_4(x)  & 5 p_5(x) &  5 p_6(x) \\
\end{array}\right )\ .
\end{equation}
Computing its largest eigenvalue and using (\ref{nearest}) one finds that  
the nearest-neighbour distribution in this case  for non-symmetric matrices has the form
\begin{eqnarray}
p(s)&=&(a_2 s^2+a_3s^3+a_4s^4+ a_5 s^5 + a_6 s^6\nonumber \\
&+&a_7 s^7 +a_8 s^8+a_9 s^9+a_{10} s^{10} )\mathrm{e}^{-7s} 
\label{ps_7_2}
\end{eqnarray}
where coefficients $a_n$ are: 
$a_2 \simeq 3.4998,\;
a_3 \simeq 24.4986,\;
a_4 \simeq 82.4309,\;
a_5 \simeq 176.8723,\;
a_6 \simeq 251.6396,\;
a_7 \simeq 229.5488,\;
a_8 \simeq 130.8981,\;
a_9 \simeq 43.7932,\;
a_{10} \simeq 6.8214.$

For symmetric matrices for the same $\alpha$ and $N$ 
\begin{eqnarray}
p(s)&=&(a_{1/2} \sqrt{s}+a_1 s+a_{3/2} s^{3/2}+a_2 s^2+ a_{5/2} s^{5/2}\nonumber\\
 &+& a_3 s^3
+ a_{7/2} s^{7/2}+a_4 s^4+a_{9/2} s^{9/2} ) {\rm e}^{-7s/2} 
\label{pss_7_2}
\end{eqnarray}
with 
$a_{1/2}\simeq .1508,\;
a_1 \simeq .7500,\;
a_{3/2} \simeq 2.0293,\;
a_2 \simeq 3.8675,\;
a_{5/2} \simeq 5.3099,\;
a_3 \simeq 5.0193,\;
a_{7/2} \simeq 3.1567,\;
a_4 \simeq 1.2312,\;
a_{9/2} \simeq 0.2350.$

\begin{figure}
 \begin{center}
\includegraphics[width=.3\linewidth]{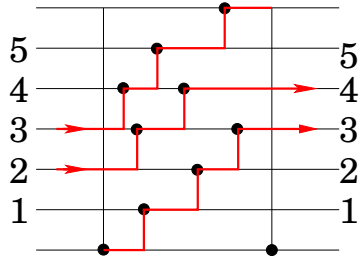}
\end{center}
\caption{One of transfer matrix elements for $\alpha=1/7$  and $N\equiv -3$ mod $7$.}
\label{m_7_23_34}
\end{figure}

For $\alpha=1/7$  and $N\equiv \pm 3$ mod $7$ there exist two possible entering lines and two leaving lines (cf. Fig.~\ref{m_7_23_34})
The dimension of the transfer matrix in this case is $C_{5}^2=10$. Its
explicit form is the following  
\begin{equation}
\hspace{-2cm} 
{\small
T(x)=\left(
\begin{array}{cccccccccc}
 10 p_6 & 35 p_7&  70 p_8&  84 p_9&  56 p_8& 168 p_9
 & 252 p_{10}& 210 p_{10} & 462 p_{11}& 462 p_{12} \\
 10 p_5 & 35 p_6 &  70 p_7 &  84 p_8 &  56 p_7 & 168 p_8 &
  252 p_9& 210 p_9 & 462 p_{10} &462 p_{11} \\
  6 p_4 & 20 p_5 &  40 p_6 &  49 p_7 &  30 p_6 &  91 p_7 &
   140 p_8 & 112 p_8  & 252 p_9 &252 p_{10} \\
  3 p_3 &  8 p_4 &  15 p_5 &  19 p_6 &  10 p_5 &  30 p_6 & 
   49 p_7 &  35 p_7  &  84 p_8  & 84 p_9 \\
  4 p_4 & 15 p_5 &  30 p_6 &  35 p_7 &  26 p_6 &  77 p_7& 
  112 p_8 &  98 p_8 & 210 p_9  &210 p_{10} \\
  3 p_3 & 12 p_4 &  25 p_5 &  30 p_6 &  20 p_5 &  61 p_6 & 
   91 p_7 &  77 p_7 &  168 p_8  &168 p_9  \\
  2 p_2 &  6 p_3 &  12 p_4 &  15 p_5 &   8 p_4 &  25 p_5 &  
  40 p_6 &  30 p_6 &  70 p_7 & 70 p_8 \\
  0 &  3 p_3 &   8 p_4 &  10 p_5 &   6 p_4 &  20 p_5 &  
  30 p_6 &  26 p_6  &  56 p_7  & 56 p_8 \\
  0 &  2 p_2 &   6 p_3 &   8 p_4 &   3 p_3 &  12 p_4 & 
   20 p_5 &  15 p_5  &  35 p_6 & 35 p_7 \\
  0 & 0 &   2 p_2 &   3 p_3 &  0 &   3 p_3 &   
  6 p_4  & 4 p_4  & 10 p_5  &10 p_6 
\end{array}\right)\ .}
\end{equation}
Finally one obtains that for $\alpha=1/7$ and $N\equiv \pm 3$ mod $7$ the
nearest-neighbour distribution for non-symmetric ensemble   is   
\begin{eqnarray}
p(s)&=&( a_2 s^2+a_3s^3+a_4s^4+ a_5 s^5 + a_6 s^6+a_7 s^7\nonumber \\
&& +a_8 s^8+a_9 s^9+a_{10} s^{10}+a_{11} s^{11}+a_{12} s^{12} ) \mathrm{e}^{-7s} 
\label{ps_7_3}
\end{eqnarray}
where
$a_2 \simeq 4.056,\;
a_3 \simeq 28.3898,\;
a_4 \simeq 91.6591,\;
a_5 \simeq 177.9134,\;
a_6 \simeq 227.8782,\;
a_7 \simeq 200.0096,\;
a_8 \simeq 121.6091,\;
a_9 \simeq 50.5880,\;
a_{10} \simeq 13.778,\;
a_{11} \simeq 2.2159,\;
a_{12} \simeq .1596.$

For symmetric matrices under the same conditions
\begin{eqnarray}
p(s)&=&(a_{1/2} \sqrt{s}+a_1 s+a_{3/2} s^{3/2}+a_2 s^2+ a_{5/2} s^{5/2} + a_3 s^3\nonumber \\
&+& a_{7/2} s^{7/2}
 +a_4 s^4+a_{9/2} s^{9/2}+a_{5} s^{5}+a_{11/2} s^{11/2} )\mathrm{e}^{-7s/2} 
\label{pss_7_3}
\end{eqnarray}
with
$a_{1/2} \simeq .1747,\;
a_1 \simeq .8691,\;
a_{3/2} \simeq 2.2565,\;
a_2 \simeq 3.8902,\;$
$a_{5/2} \simeq 4.8085,\;
a_3 \simeq 4.3734,\;
a_{7/2} \simeq 2.9327,\;
a_4 \simeq 1.4222,\;$
$a_{9/2} \simeq .4747,\;
a_{5} \simeq .0979,\;
a_{11/2} \simeq .0094.$

\begin{figure}
\begin{minipage}{.49\linewidth}
\begin{center}
\includegraphics[angle=-90, width=.9\linewidth]{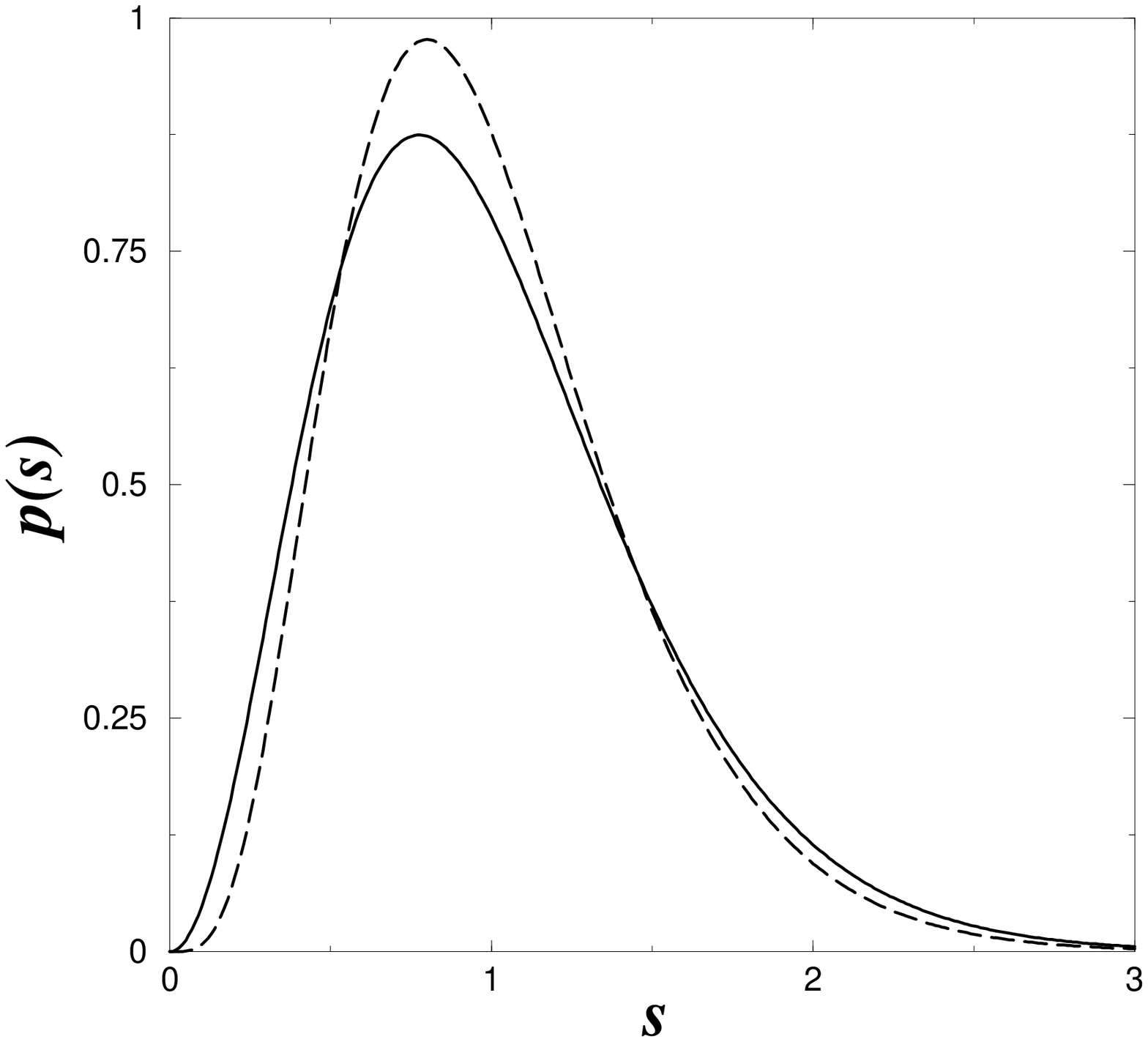}
\end{center}
\begin{center}a)\end{center}
\end{minipage}\hfill
\begin{minipage}{.49\linewidth}
\begin{center}
\includegraphics[angle=-90, width=.9\linewidth]{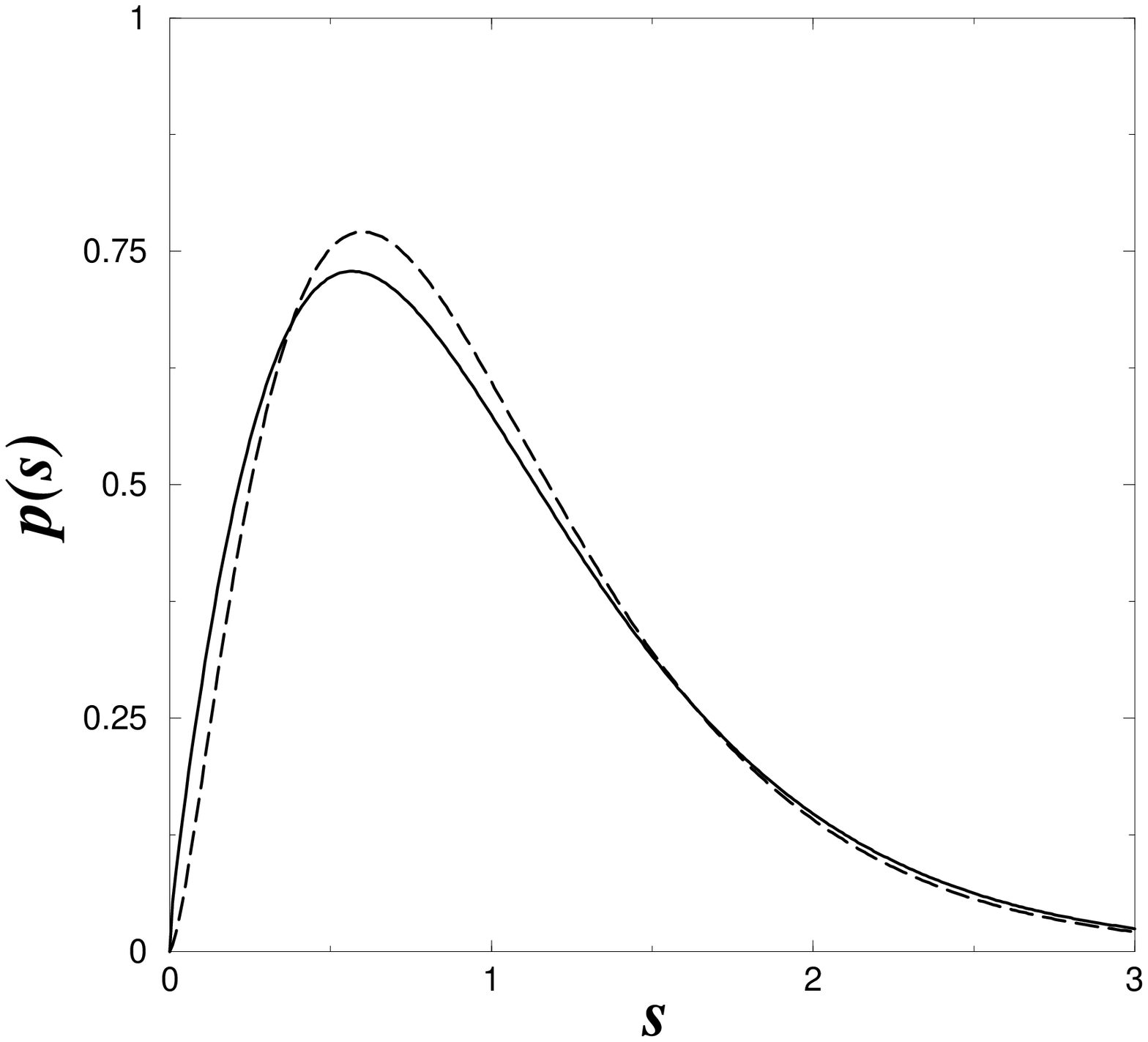}
\end{center}
\begin{center}b)\end{center}
\end{minipage}
\caption{Nearest-neighbour distribution for $\alpha=1/5$ for (a) the
  non-symmetric ensemble and (b) the symmetric one. Dashed lines correspond to
  $N\equiv \pm 1$ mod $5$ given by (\ref{nnd}) with $\beta=4$ in (a) and
  $\beta=3/2$ in (b). Solid lines indicate the results for  $N\equiv \pm 2$
  mod $5$ given by (\ref{ps_5}) in (a) and by (\ref{pss_5}) in (b). } 
\label{fig_5_5s}
\end{figure}

\begin{figure}
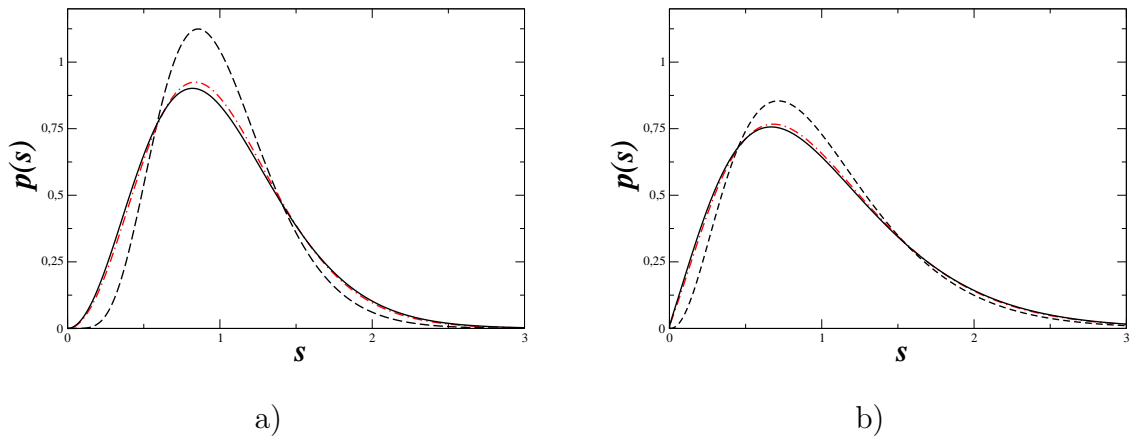

\vspace{0.5cm}
\begin{minipage}{.49\linewidth}
\begin{center}
\includegraphics[width=.9\linewidth]{fig11a.eps}
\end{center}
\begin{center}a)\end{center}
\end{minipage}\hfill
\begin{minipage}{.49\linewidth}
\begin{center}
\includegraphics[width=.9\linewidth]{fig11b.eps}
\end{center}
\begin{center}b)\end{center}
\end{minipage}
\caption{The same as in Fig.~\ref{fig_5_5s} but for $\alpha=1/7$.  Dashed
  black lines correspond to $N\equiv \pm 1$ mod $7$ given by (\ref{nnd}) with
  $\beta=6$ in non-symmetric matrices and $\beta=3/2$ for symmetric
  matrices. The red dotted dashed lines indicate the results for $N\equiv \pm
  2$ mod $7$ 
  given by (\ref{ps_7_2}) in (a) and by (\ref{pss_7_2}) in (b). Solid black
  lines represent the results for  $N\equiv \pm 3$ mod $7$ determined by
  (\ref{ps_7_3}) in (a) and by (\ref{pss_7_3}) in (b). } 
\label{fig_7_7s}
\end{figure}

In Fig.~\ref{fig_5_5s} and \ref{fig_7_7s}  the calculated nearest neighbour
distributions are plotted for $\alpha=1/5$ and $\alpha=1/7$ with all possible
residues of $N\not \equiv 0$ modulo $1/ \alpha$. As expected, the case
$N\equiv \pm 1$ mod $1/ \alpha$ differs considerably from other cases. When
the residue, $k$, increases  the nearest-neighbour distribution more and more
resembles to the nearest-neighbour distribution of the standard Gaussian
ensembles of random matrices. For example, for $\alpha=1/7$ the results with
$N\equiv \pm 2$ and    $N\equiv \pm 3$ are difficult to distinguish  from the
Wigner surmise (\ref{surmise})  for GUE (for non-symmetric matrices) and for
GOE (for symmetric ones)	 

\begin{figure}
\begin{minipage}{.49\linewidth}
\begin{center}
\includegraphics[angle=-90, width=.9\linewidth]{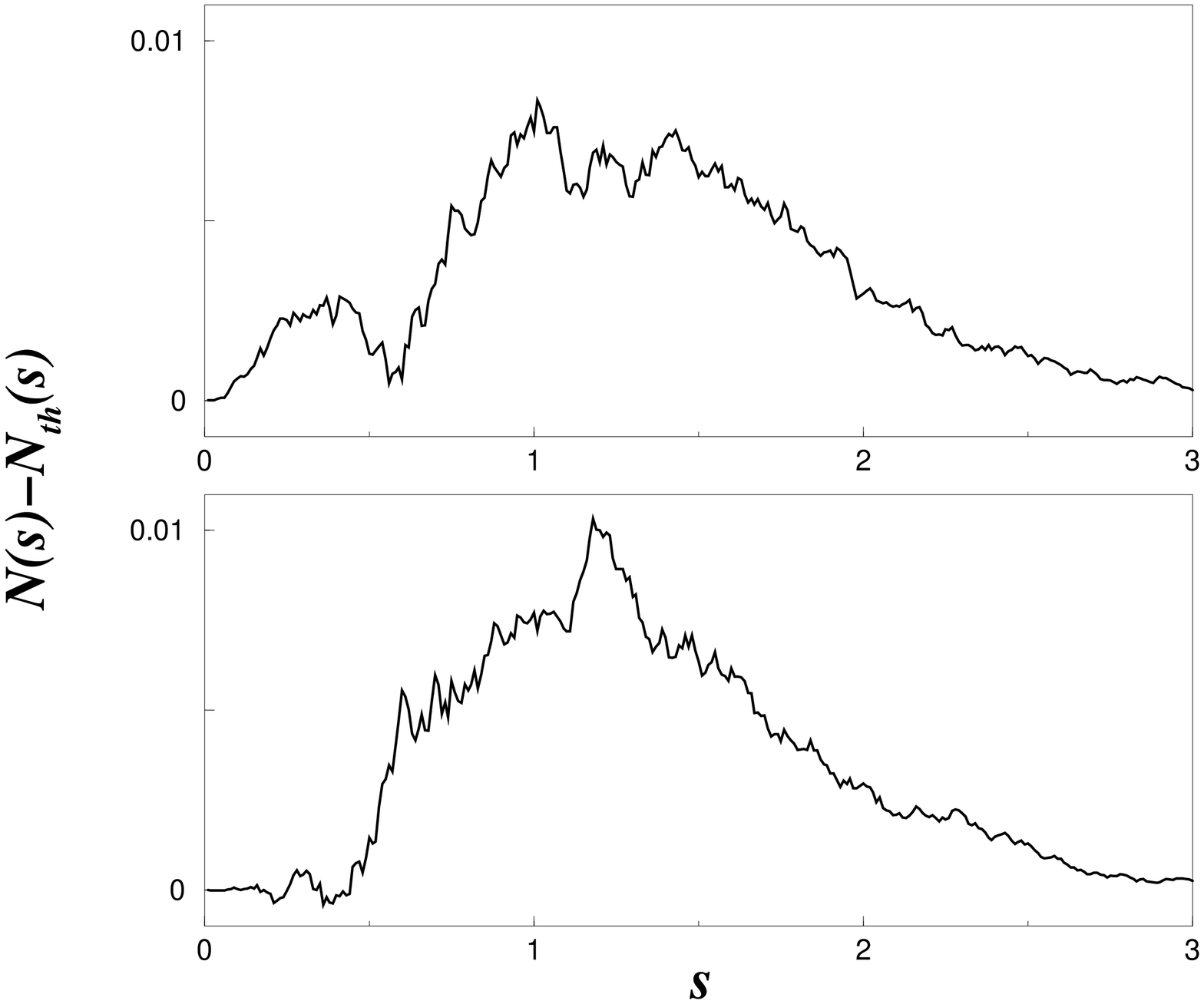}
\end{center}
\begin{center}a)\end{center}
\end{minipage}\hfill
\begin{minipage}{.49\linewidth}
\begin{center}
\includegraphics[angle=-90, width=.9\linewidth]{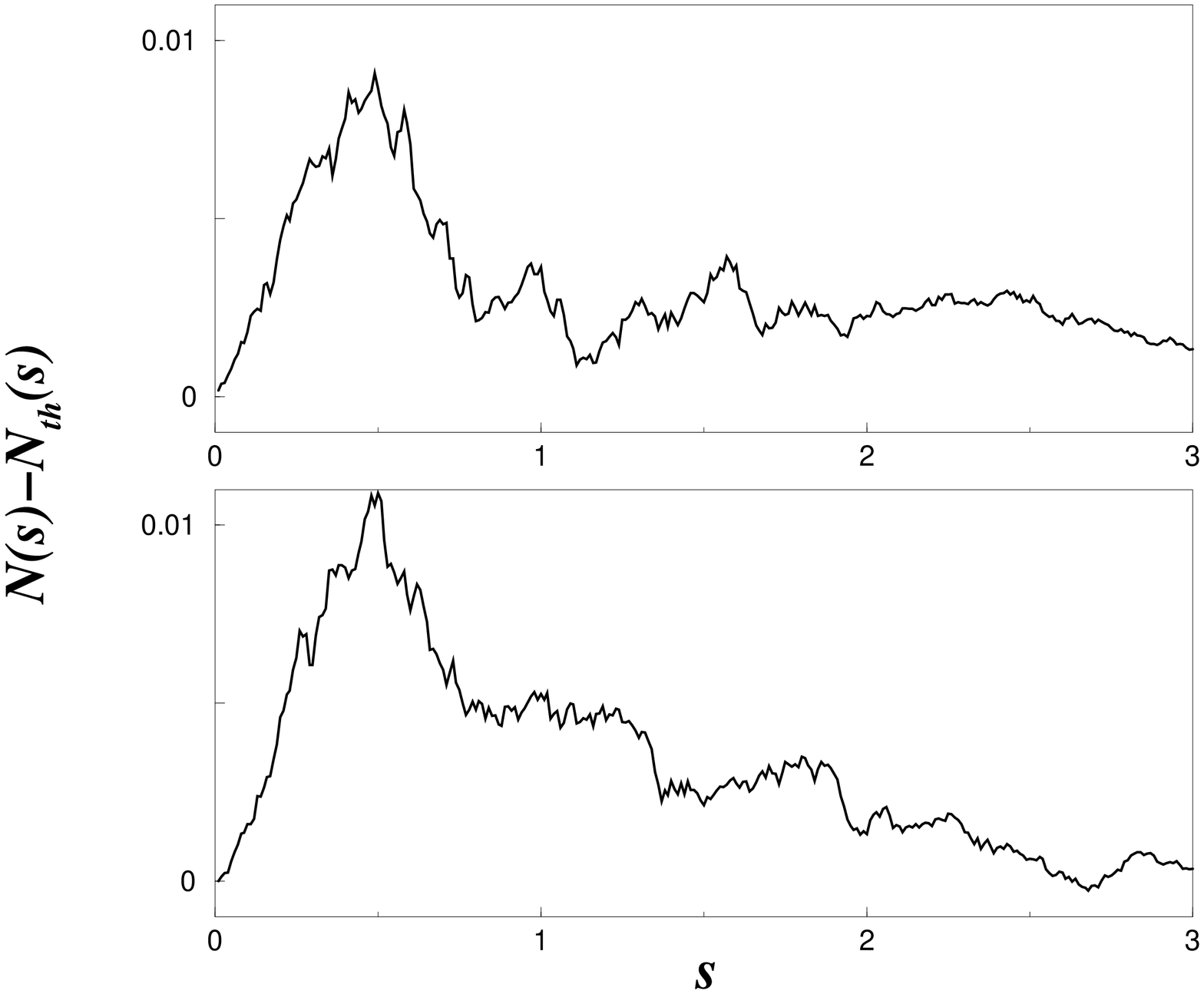}
\end{center}
\begin{center}b)\end{center}
\end{minipage}
\caption{Difference between the integrated nearest-neighbour distribution and
  the corresponding theoretical prediction for $\alpha=1/5$. (a) For
  non-symmetric ensembles. (b) For symmetric matrices. In each graph pictures
  differ by the matrix dimensions. From bottom to top $N=801$ ($N\equiv 1$ mod
  $5$) and $802$ ($N\equiv 2$ mod $5$). } 
\label{fig_N_5}
\end{figure}

\begin{figure}
\begin{minipage}{.49\linewidth}
\begin{center}
\includegraphics[angle=-90, width=.9\linewidth]{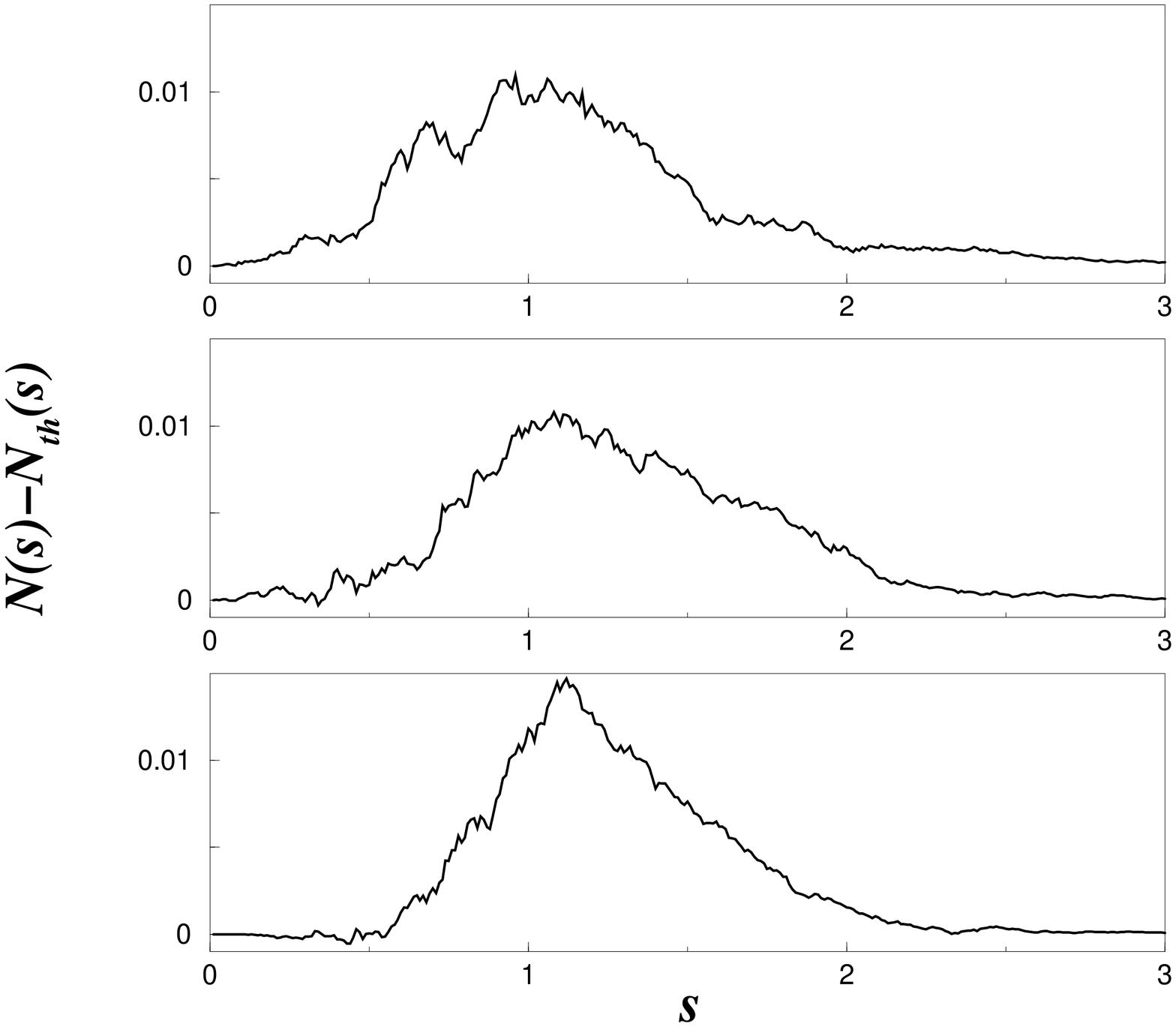}
\end{center}
\begin{center}a)\end{center}
\end{minipage}\hfill
\begin{minipage}{.49\linewidth}
\begin{center}
\includegraphics[angle=-90, width=.9\linewidth]{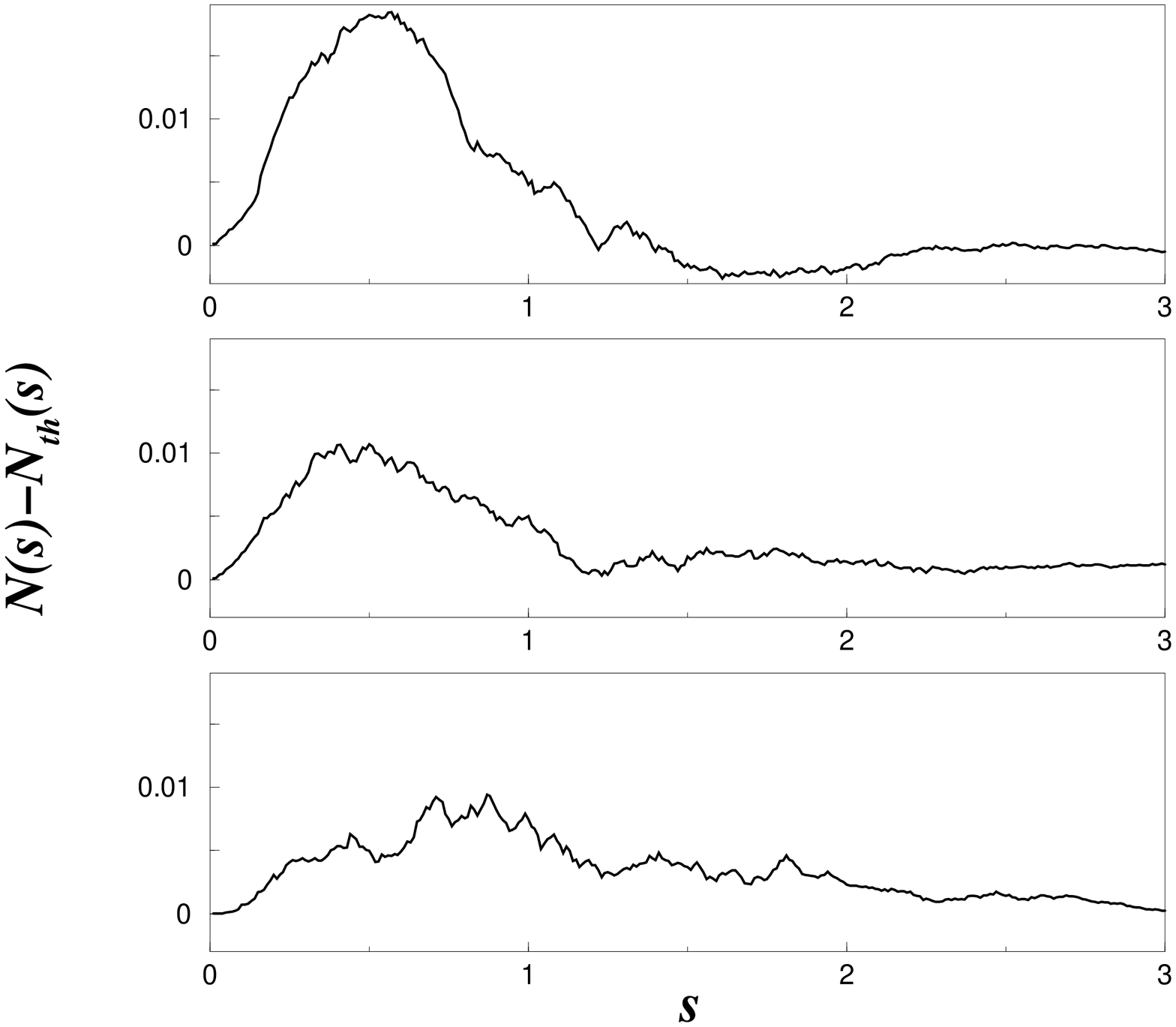}
\end{center}
\begin{center}b)\end{center}
\end{minipage}
\caption{The same as in Fig.~\ref{fig_N_5} but for $\alpha=1/7$. From bottom
  to top $N=799$ ($N\equiv 1$ mod $7$), $800$ ($N\equiv 2$ mod $7$)  , and
  $801$ ($N\equiv 3$ mod $7$). } 
\label{fig_N_7}
\end{figure}

To compare these formulas with the results of numerical simulations it is more
precise to use the integrated nearest-neighbour distribution
(\ref{integrated_nnd}). In Figs.~\ref{fig_N_5} and \ref{fig_N_7} such
comparison is performed for all cases considered. The agreement is quite good
and the differences are of the same order as in Fig.~\ref{fig0}.

\section{Summary}\label{summary}

A unitary random matrix ensemble 
\begin{equation}
 M_{kp}=\mathrm{e}^{{\mathrm{i}}\Phi_k}\frac{(1-{\rm e}^{2\pi {\rm i}\alpha N})}{N(1-{\rm e}^{2\pi {\rm i}(k-p+\alpha N)/N})}
\end{equation}
corresponding to a quantization of a simple pseudo-integrable map (\ref{map})
is considered in detail. These matrices are  characterized by a rational
parameter 
$\alpha=m/q$, the matrix dimension $N$, and symmetry properties of random
phases $\Phi_k$ (\ref{symmetry}). To get a well defined limit of the spectral
statistics of these ensembles for large $N$ it is necessary to consider
increasing sequences of $N$ such that $mN$ has a fixed residue modulo the
denominator of $\alpha$ 
\begin{equation}
mN\equiv k \;\mathrm{mod}\;q
\end{equation}
with the residue $k=0,1,\ldots,q-1$. For $k=0$  all eigenvalues of the main
matrix can be found analytically as in \cite{zeev}.  For all other residues
the spectral statistics of the considered ensembles is non-trivial and differs
considerably from standard random matrix ensembles. The cases $k=1$ and
$k=q-1$  have been investigated in \cite{Schmit} where it was shown that for
these $k$ the nearest-neighbour distribution has the following form   
\begin{equation}
p(s)\sim \left  \{\begin{array}{ll}s^{q-1}\mathrm{e}^{-qs}&\mathrm{for\; non-symmetric\; ensemble }\\
s^{q/2-1}\mathrm{e}^{-qs/2}&\mathrm{for\; symmetric\; ensemble}\end{array}\right . .
\label{results}
\end{equation}
In the present paper a kind of transfer operator method is developed  to
calculate the spectral statistics of the same matrix for all values of $k$. It
is demonstrated that  
the nearest-neighbour distribution equals the product of a finite polynomial in
$s$ for non-symmetric matrices and in $\sqrt{s}$ for symmetric matrices times
the same exponential factor as in (\ref{results}) 
\begin{equation}
p(s)= \left  \{\begin{array}{ll} \displaystyle\sum_{n=2}^{k(q-k)}a_n s^n
    \mathrm{e}^{-qs}&\mathrm{for\; non-symmetric\; ensemble }\\  
\displaystyle\sum_{n=1}^{(k(q-k)-1)}a_{n/2} s^{n/2}
\mathrm{e}^{-qs/2}&\mathrm{for\; symmetric\; ensemble}\end{array}\right . . 
\end{equation} 
Statistical properties of sequences with residue $k$ and $q-k$ are the same.
The values of coefficients $a_n$ can be calculated by finding the largest
eigenvalue and corresponding left and right eigenvectors of the transfer
matrix.  

For  $\alpha=1/5$ and $\alpha=1/7$ and all possible residues the explicit form
of these coefficients have been calculated. Numerical simulations in these
cases are in a good agreement with  obtained formulas. Other correlation
functions can also be expressed from the same quantities.   

It appears that the considered ensembles of random matrices permit
different generalizations which will be discussed elsewhere
\cite{BogomolnyGiraud}. 

\section*{Acknowledgments}

It is a pleasure to thank O. Bohigas and O. Giraud for helpful discussions and
J. Marklof for a careful reading of the manuscript. RD
wishes to acknowledge financial support from the ``Programme Lavoisier'' of 
the French Minist\`ere des Affaires \'etrang\`eres et europ\'eennes and EPSRC.  

\appendix
\section{} \label{appendix}

The purpose of the Appendix is to give the proofs of certain formulas used in the text.

Let for all $n=1,\ldots, N$ 
\begin{equation}
 \sum_{m=1}^N\frac{b_m}{x_m-y_n}=1\;.
\end{equation}
Solutions $b_m$ of these equations can be expressed in terms of Cauchy determinants and 
\begin{equation}
 b_m=\frac{\displaystyle\prod_n(x_m-y_n)}{\displaystyle\prod_{s\neq
     m}(x_m-x_s)} \;.
\label{bm}
\end{equation}
A simple way to check it is to consider the function
\begin{equation}
 f_n(x)=\frac{\displaystyle\prod_{r\neq
     n}(x-y_r)}{\displaystyle\prod_{s}(x-x_s)}
 =\frac{\displaystyle\prod_{r}(x-y_r)}{(x-y_n)\displaystyle\prod_{s}(x-x_s)}\;. 
\end{equation}
Asymptotically $f_n(x)\to 1/x$ so the integral over a big contour encircling
all poles equals 1. Rewriting this integral as the sum poles gives 
\begin{equation}
 1=\sum_m\frac{\displaystyle\prod_r(x_m-y_r)}{(x_m-y_n)
   \displaystyle\prod_{s\neq m}(x_m-x_s)} 
\end{equation}
which proves (\ref{bm}).

Denote 
\begin{equation}
 g(x)=\frac{\displaystyle\prod_n(x-\Lambda_n{\rm e}^{2\pi {\rm
       i}\alpha})}{x\displaystyle\prod_{k}(x-\Lambda_k)}\;. 
\end{equation}
This function decreases as $1/x$ for large $x$ and has poles at $x=0$ and
$x=\Lambda_k$. Integrating it over a contour encircling all poles one gets 
\begin{equation}
 1={\rm e}^{2\pi {\rm
     i}\alpha}+\sum_{m=1}^N\frac{\displaystyle\prod_n(\Lambda_m- \Lambda_n{\rm
     e}^{2\pi {\rm i}\alpha})} 
{\Lambda_m \displaystyle\prod_{k\neq m}(\Lambda_m-\Lambda_k)}
\end{equation}
from which it follows that $|B_m|^2$ defined by (\ref{coefficients}) obey
automatically the normalization condition (\ref{norm}).

\end{document}